\documentclass{aa}  

\usepackage{graphicx}
\usepackage{enumerate}
\usepackage{natbib}
\bibpunct{(}{)}{;}{a}{}{,}
\usepackage[varg]{txfonts}
\usepackage[utf8]{inputenc}

\usepackage{multirow}
\usepackage{makecell}
\usepackage{amsmath}
\usepackage[colorlinks=true,urlcolor=blue,citecolor=blue,linkcolor=blue]{hyperref}

\begin{document}

	\title{The solar chromosphere at millimetre and ultraviolet wavelengths}
	\subtitle{I. Radiation temperatures and a detailed comparison}

   \author{S.~Jafarzadeh\inst{1,2}, S.~Wedemeyer\inst{1,2}, M.~Szydlarski\inst{1,2}, B.~De~Pontieu\inst{3,1,2}, R.~Rezaei\inst{4,5}, \and M. Carlsson\inst{1,2}}

   \institute{Rosseland Centre for Solar Physics, University of Oslo, P.O. Box 1029 Blindern, N-0315 Oslo, Norway\\ 
   		\email{shahin.jafarzadeh@astro.uio.no}
   		\and
   		Institute of Theoretical Astrophysics, University of Oslo, P.O. Box 1029 Blindern, N-0315 Oslo, Norway
   		\and
   		Lockheed Martin Solar \& Astrophysics Lab, Org. A021S, Bldg. 252, 3251 Hanover Street Palo Alto, CA 94304, USA
   		\and
   		Instituto de Astrof\'{i}sica de Canarias, Avda V\'{i}a L\'{a}ctea S/N, La Laguna 38200, Tenerife, Spain
   		\and
   		Departamento de Astrof\'{i}sica, Universidad de La Laguna, 38205 La Laguna (Tenerife), Spain\\             
             }

   \date{Received 7 September 2018 / Accepted 24 December 2018}

  \abstract{Solar observations with the Atacama Large Millimeter/submillimeter Array (ALMA) provide us with direct measurements of the brightness temperature in the solar chromosphere. We study the temperature distributions obtained with ALMA Band~6 (in four sub-bands at 1.21, 1.22, 1.29, and 1.3~mm) for various areas at, and in the vicinity of, a sunspot, comprising quasi-quiet and active regions with different amounts of underlying magnetic fields. We compare these temperatures with those obtained at near- and far-ultraviolet (UV) wavelengths (and with the line-core intensities of the optically-thin far-UV spectra), co-observed with the Interface Region Imaging Spectrograph (IRIS) explorer. These include the emission peaks and cores of the Mg\,{\sc ii}\,k 279.6~nm and Mg\,{\sc ii}\,h 280.4~nm lines as well as the line cores of C\,{\sc ii}~133.4~nm, O\,{\sc i}~135.6~nm, and Si\,{\sc iv}~139.4~nm, sampling the mid-to-high chromosphere and the low transition region.
  Splitting the ALMA sub-bands resulted in an slight increase of spatial resolution in individual temperature maps, thus, resolving smaller-scale structures compared to those produced with the standard averaging routines.
  We find that the radiation temperatures have different, though somewhat overlapping, distributions in different wavelengths and in the various magnetic regions. Comparison of the ALMA temperatures with those of the UV diagnostics should, however, be interpreted with great caution, the former is formed under the local thermodynamic equilibrium (LTE) conditions, the latter under non-LTE. The mean radiation temperature of the ALMA Band~6 is similar to that extracted from the IRIS C\,{\sc ii} line in all areas with exception of the sunspot and pores where the C\,{\sc ii} poses higher radiation temperatures. In all magnetic regions, the Mg\,{\sc ii} lines associate with the lowest mean radiation temperatures in our sample. These will provide constraints for future numerical models.
  }

   \keywords{Sun: chromosphere --
             Sun: transition region -- 
             Sun: radio radiation --
             Sun: UV radiation
               }

  \authorrunning{S.~Jafarzadeh~et~al.}
  \titlerunning{The solar chromosphere at millimetre and ultraviolet wavelengths}
   \maketitle

\section{Introduction}
\label{sec-intro}

The highly structured chromosphere of the Sun exhibits high temperatures 
implying that this layer is heated. Various heating mechanisms have been proposed \citep[e.g.][]{Cram1977,Kalkofen2007,Carlsson2010}. The magnetic field, in particular, is shown to play a crucial role in the dynamics and energetics of the chromosphere (and of the transition region), and thus affects its temperature distribution \citep[e.g.][]{Withbroe1977,Judge1998,Rutten2007,Hasan2008,de-la-Cruz-Rodriguez2013}. 
Among different mechanisms, two magnetic processes have mainly been favoured in this context: magnetic reconnection (Joule heating; \citealt{Parker1988}) and propagation of magneto-acoustic (or magneto-hydrodynamic; MHD) waves (mechanical heating; \citealt{Osterbrock1961,Rabin1984}). The latter have been shown to carry a large enough energy flux to, in principle, heat the chromosphere, but it remains unclear whether their dissipation contributes significantly to the local energy balance (e.g. \citealt{DePontieu2007,Khomenko2008,De-Pontieu2012,Morton2013,Gafeira2017,Jafarzadeh2017a,Jafarzadeh2017b}; see \citealt{Jess2015} and \citealt{Khomenko2015} for recent reviews).

Using numerical simulations, \citet{Carlsson1995} showed that the rise of the gas temperature with height in the quiet-Sun chromosphere is different compared to the rise of the radiation temperature. The former changes very little, whereas the latter shows a chromospheric rise similar to the standard models, such as the FAL model atmospheres \citep{Fontenla1993}.
Moreover, it has been shown that there is not always a one-to-one relationship between gas temperatures ($T_{gas}$) and radiation temperatures ($T_{rad}$) of UV channels (i.e. the source functions of the UV lines are not always coupled to the Planck function). For instance, \citet{Leenaarts2013b} shows that the source function of emission peaks of the Mg\,{\sc ii} lines (forming in the mid-chromosphere) is partially decoupled from the local temperature, thus they only find a good $T_{gas}-T_{rad}$ correlation for larger intensities. For a far-UV spectral line (i.e. C\,{\sc ii}~133.4~nm), \citet{Rathore2015a} found a $T_{gas}=2\,T_{rad}$ relationship (from radiation MHD simulations).

The radiation temperature at UV wavelengths can be determined by the so-called radiometry of the emergent intensity, where calibrated data numbers are converted to intensity, and hence to radiation temperature by solving the Planck function. 

Despite the UV observations of the solar chromosphere in the past decades \citep[e.g.][]{Burton1967,de-Wijn2005,Rutten2012}, 
substantial progress has been made recently thanks to near and far UV observations at high spectral and high spatial resolution with the Interface Region Imaging Spectrograph  (IRIS; \citealt{De-Pontieu2014}) explorer. Thus, the radiation temperature in the solar chromosphere can be approximated to a higher accuracy by, for example, resolving relatively small-scale structures, which improves the derived chromospheric temperature distributions.

Observations at millimetre wavelengths are complementary to the UV diagnostics in the sense that the source function varies linearly with temperature.  
With the millimetre wavelengths being in the Rayleigh-Jeans limit,
the temperature can be simply measured under the assumption of local thermodynamic equilibrium (LTE), in contrast to UV diagnostics for which deviations from LTE cannot be neglected and thus significantly complicate the computation and interpretation of such measurements. Although a non-thermal contribution to the millimetre radiation can lead to an overestimation of the temperature in the chromosphere \citep{Shklovskii1958}, its contribution is negligible apart from during flares.

Measuring the brightness temperature at millimetre and sub-millimetre wavelengths has a long history (e.g. \citealt{Kundu1965,Kundu1982,White1992}, and references therein). However, older studies have been limited to a very low spatial resolution, mainly due to the fact that they were based on single-dish measurements (see discussion in \citealt{White2017}).
Yet,  interferometric observations at millimetre and sub-millimetre wavelengths have been challenging in the past, mainly due to effects from the Earth’s atmosphere for observations from low-altitude sites \citep{White2006,Loukitcheva2006}.
Only recently, the Atacama Large Millimeter/submillimeter Array (ALMA; \citealt{Wootten2009,Hills2010}) provides a leap forward for this type of chromospheric diagnostics \citep{Wedemeyer2016a,Wedemeyer2016b}.

For an optically thick line, the radiation temperature gives information on the source function around optical depth unity. In LTE, this translates directly to the gas temperature while in non-LTE the source function is typically more or less decoupled and typically lower than the Planck function. For an optically thin line, the radiation temperature is not a meaningful quantity since the intensity is given by an integral of the emissivity along the line of sight and is not given by the source function.

Agreements between the gas temperature deduced from the brightness temperature of the UV and millimetre observations in the solar chromosphere have been previously reported using low resolution observations and model atmospheres \citep{Cuny1971,Vernazza1981}. Such agreements, to some degree, have been also provided for the millimetre and UV observations with ALMA and IRIS by \citet{Bastian2017,Bastian2018}, who compared the brightness temperature of the ALMA's band 6 at 1.25~mm and the radiation temperature of the average of the emission peaks of the IRIS Mg\,{\sc ii}~h.

The present work provides distributions of the radiation temperature at various chromospheric heights sampled by the ALMA 1.3~mm passband and the IRIS observations at near- and far-UV wavelengths (Sects.~\ref{sec-obs} and \ref{sec-analys}). The comparison between the different distributions are performed for various solar regions with different amounts of underlying magnetic fields. 
In addition, intensities of the optically thin lines in the IRIS far-UV range have been compared with those of the millimetre observations. 
The concluding remarks of our study are addressed in Sect.~\ref{sec-conclusion}.

\section{Observations}
\label{sec-obs}

This study is primarily based on observations of the solar active region AR~12470 by ALMA on 18 December 2015. These have been complimented by co-observations of the same target with the IRIS explorer and with the Solar Dynamic Observatory (SDO; \citealt{Pesnell2012}). All images from IRIS and SDO are precisely co-aligned with that from ALMA (i.e. the IRIS and SDO images are first re-scaled to the ALMA images, then they are transformed (spatially shifted and/or rotated) based on cross-correlation of common features). The images from the different instruments are analysed with their original spatial resolution (but they are all re-sampled to match the ALMA's image scale) when distributions of radiation temperature are determined (see Section~\ref{subsec-histograms}). 
The exception is when the one-to-one comparison between radiation temperatures (and intensities) of ALMA and IRIS is performed (see Sect.~\ref{subsec-scatterplots}). In the latter case, the IRIS raster images are convolved with the point spread function (PSF) of the ALMA map.

\begin{figure*}[!thp]
\centering
    \includegraphics[width=.87\textwidth, trim = 0 0 0 0, clip]{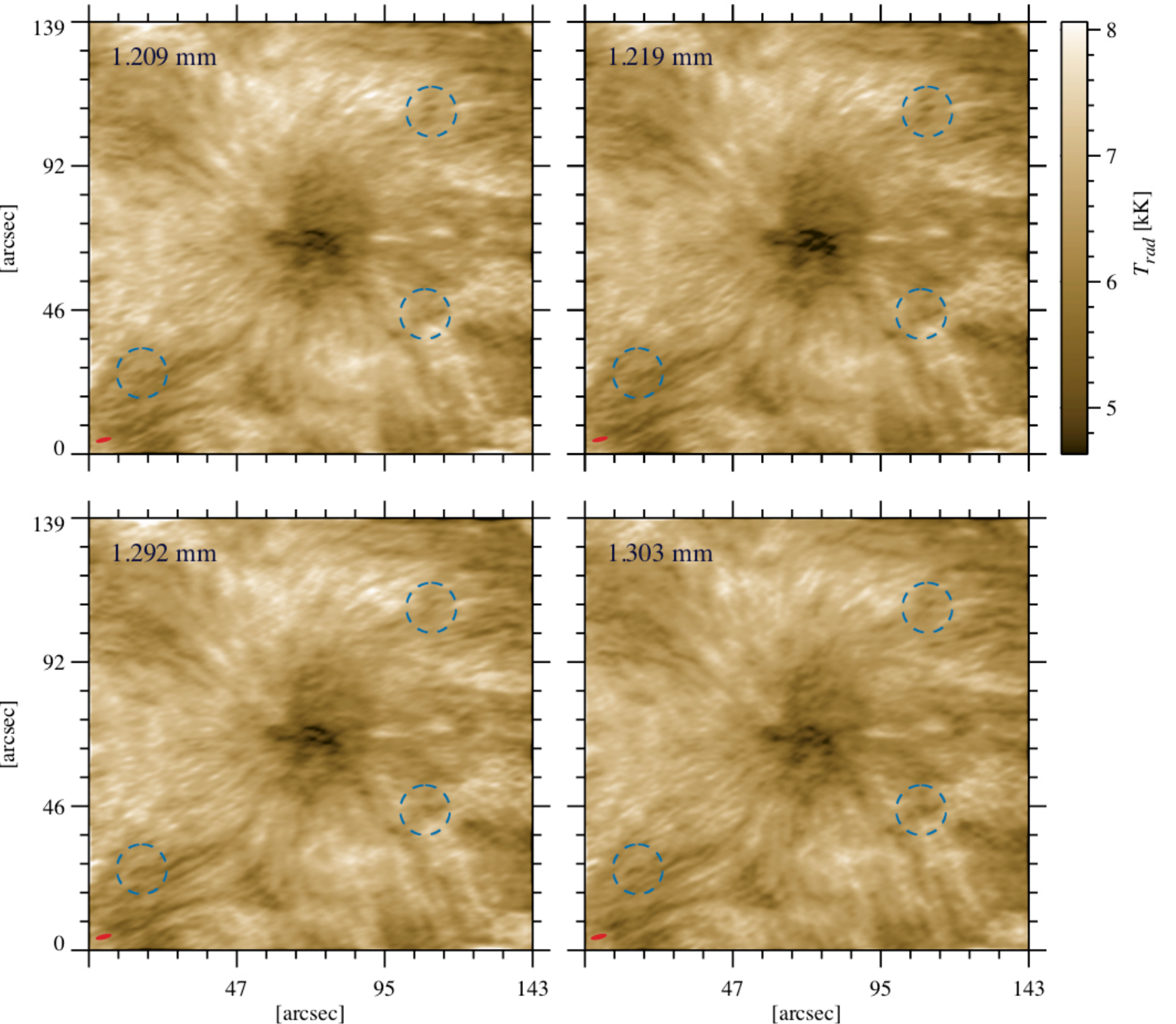}
  \caption{ALMA sub-band images of brightness temperature in band 6. The sampling wavelengths are indicated in the upper-left corner of each panel. The circles mark examples of structures with different temperature distributions in the sub-band images, sampling various (but close) heights in the solar chromosphere. The red ellipses in the lower-left corners illustrate the beam size (i.e. size of the resolution element) of each sub band.}
  \label{fig:almasubbands}
\end{figure*}

\subsection{Observations with ALMA}
\label{subsec-almaobs}

A mosaic image with 149~points, covering an area of ($\approx140\times140$)~arcsec$^2$ on the Sun (with a sampling resolution of $\approx0.15$~arcsec/pixel), was obtained in ALMA band 6 (centred at 239~GHz; 1.25~mm) on 18 December 2015 between 19:39:25--20:02:34~UT (in the 6th solar commissioning science verification campaign). 
The mosaic observations were performed with a pointing separation of 11.2~arcsec (corresponding to Nyquist sampling) and an integration time of 6.048~s for each point.
The field of view (FOV) includes the leading sunspot of the active region AR~12470 surrounded by plages and quasi-quiet\footnote[1]{The somewhat quiet area within the immediate vicinity of the active region, in the photosphere. Thus, it is largely influenced by the magnetic canopy at the chromospheric heights and beyond.} areas, close to the solar disc-center (with cosine of heliocentric angle $\approx0.97$).

The data was taken using a heterogeneous array configuration consisting of twenty-one 12~m and nine 7~m antennas. Four spectral windows (sub bands), with a bandwidth of 2~GHz each, were recorded. Each spectral window (SW) includes 128 channels (with time domain mode (TDM) as the spectral setup). 
In addition, full-disc total power maps were simultaneously obtained through  single dish observations (i.e. with the total power (TP)-array in fast-scanning mode).  
The TP maps are used to derive absolute brightness temperatures for the interferometric data  \citep{Shimojo2017,White2017}. 
We have optimised the standard reduction pipeline by tuning various parameters, also separating the sub-bands and the spectral channels (instead of a spectral integration, which is done by default in the standard pipeline). The first version of this Solar ALMA Pipeline (SoAP) will be described in a forthcoming publication \citep{Szydlarski2018}.

\begin{figure}[!hpt]
    \centering
    \includegraphics[width=9.0cm, trim = 0 0 0 0, clip]{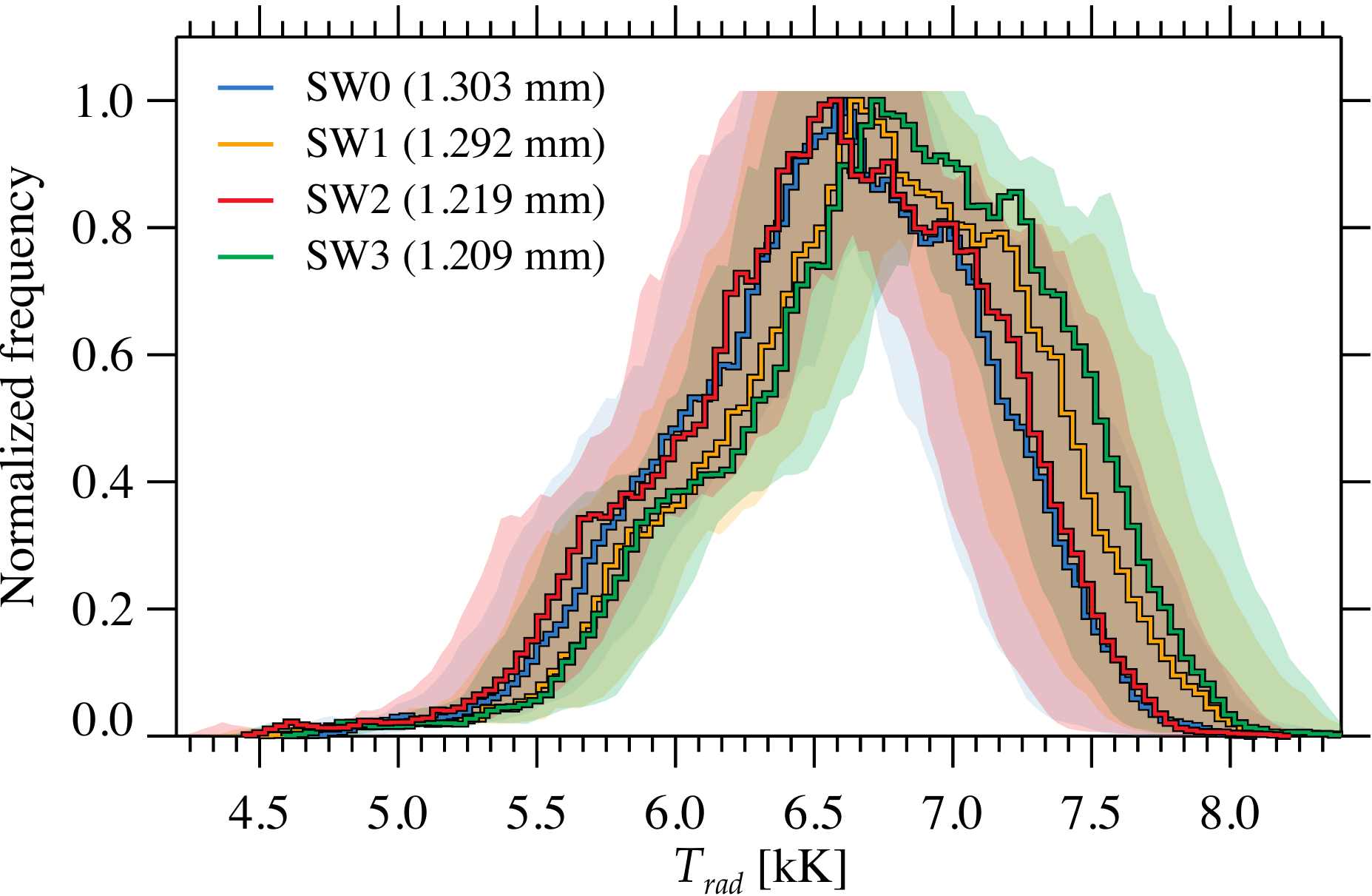}
    \caption{Distributions of brightness temperature of the entire field-of-view of the ALMA sub-band images shown in Figure~\ref{fig:almasubbands}. The histograms are normalized to their maximum occurrence frequency. The shaded areas mark the 5\% measurements' uncertainty.}
  \label{fig:almasubbandsstat}
\end{figure}

Extraction of the four different spectral windows (i.e. with ALMA identity numbers 5,7,9, and 11) included standard CASA calibrations, imaging, and feathering steps (see \citealt{Shimojo2017} for details). Thus, brightness temperature maps of the four sub bands (i.e. SW3, SW2, SW1, and SW0), corresponding to wavelengths 1.209, 1.219, 1.292, and 1.303~mm, respectively, were obtained. These correspond to frequencies 248, 246, 232, and 230~GHz, respectively. Figure~\ref{fig:almasubbands} illustrates the four sub bands. Small differences between the sampling wavelengths (hence, close heights of formation) resulted in a very similar appearances of the images. However, they represent slightly different heights of formation, thus, averaging these sub-bands (as has been commonly done), results in losing the height information and can lead to spatial smearing (i.e. reduced spatial resolution). The circles in the four panels of Figure~\ref{fig:almasubbands} serve to guide the eye to a few examples where structures have different appearances in the different sub bands. We note that the beam size, which represents the spatial resolution, slightly varies between the four sub-bands, by about 7\% from Sw0 to Sw3 (while the first- and last-two sub bands have much smaller differences in beam size). The structures marked with the circles in Figure~\ref{fig:almasubbands} are in the order of, or larger than, the beam sizes (indicated with the red ellipses in the lower-left corner of each panel in Figure~\ref{fig:almasubbands}). We therefore argue that the different appearances of the small-scale structures are mostly due to the slightly different formation heights for the sub bands although the small differences in the spatial resolution of the four sub-bands may contribute to some extent for features with sizes close to the beam sizes.

Distributions of brightness temperature of the four sub-bands of ALMA band 6 (for the entire FOV) are shown in Figure~\ref{fig:almasubbandsstat}. The shaded areas around the histograms indicate the approximate 5\% uncertainty in measurements of the brightness temperatures. A detailed discussion on the error of the temperature measurements by ALMA and of the calibrations (and to what degree the initial 5\% estimation is realistic) are subject of a forthcoming paper. The small differences of the histograms lie within the measurement error of the brightness temperature by ALMA, thus no detailed comparison of the sub bands is provided. However, it is clear that separating the sub bands (along with other improvement in the data reduction procedures) has resulted in resolving finer structures, compared to the same data reported earlier in the literature \citep[e.g.][]{Bastian2017,Shimojo2017,Loukitcheva2017}. Resolving  small-scale structures can improve the temperature distributions (otherwise spatial smearing influences the distributions; \citealt{Wedemeyer2004}). We note that although the 5\% uncertainty in measuring the brightness temperatures could also play a role in the different appearances of the small-scale structures, the spatial coherence of the features in various parts of the FOV implies that the differences are caused by real differences in the structures at the height ranges mapped by the sub-bands.

\begin{figure*}[!thp]
\centering
    \includegraphics[width=.92\textwidth, trim = 0 0 0 0, clip]{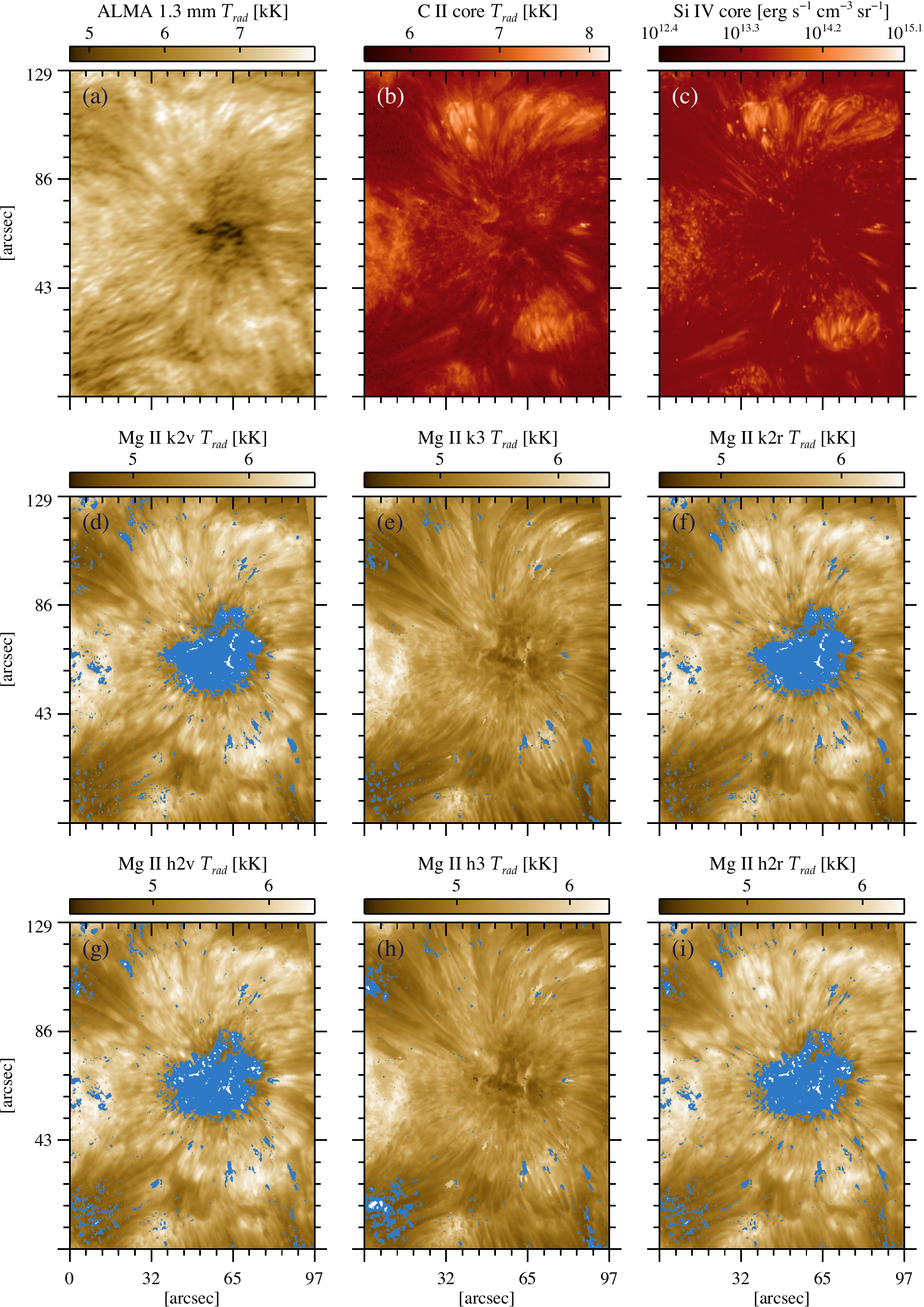}
  \caption{Radiation temperature maps of the ALMA 1.3~mm (a) and of a few IRIS line features at UV wavelengths: line core of C\,{\sc ii} 133.4~nm (c); blue emission peak, line core, and red emission peak of Mg\,{\sc ii}\,k 279.6~nm (d--f) and of Mg\,{\sc ii}\,h 280.4~nm (g--i). For comparison, line-core intensity of the IRIS Si\,{\sc iv} 139.4~nm is illustrated in panels (c). Blue pixels are regions where the corresponding line features could not be determined.
  }
  \label{fig:obs}
\end{figure*}

Among the four sub-bands, we use, in the following, the brightness temperature at 1.3~mm which supposedly corresponds to the highest sampled layer in the solar atmosphere. The ALMA 1.3~mm image is shown in the top-left panel of Figure~\ref{fig:obs}, cropped to the same FOV of the IRIS images.

\subsection{IRIS observations}
\label{subsec-irisobs}

The ALMA observations were supported by the IRIS spacecraft in the ``very large dense 320-step raster'' mode with 2~s exposure time per step. This comprises an FOV of ($105\times175$)~arcsec$^2$ (with the sampling resolution of $0.33$~arcsec/pixel along the slit and the step size of $0.35$~arcsec) overlapping a large fraction of the solar region observed by ALMA. Three full IRIS scans were recorded between 19:33:14--20:24:20 UT, of which, the first two are the closest in time to the ALMA observations (with the second one having the longest time overlap with ALMA).

We note that the IRIS raster images were spatially and spectrally binned by two pixels (on board the satellite), resulting in a higher signal-to-noise ratio (S/N). The IRIS spectrograph covers a wide wavelength range in the UV, which particularly includes the C\,{\sc ii} 133.4~nm doublet, O\,{\sc i}~135.6~nm, the Si\,{\sc iv} doublet (at 139.4~nm and 140.3~nm), Mg\,{\sc ii}\,k 279.6~nm, and Mg\,{\sc ii}\,h 280.4~nm. Among which, the magnesium lines are typically formed under optically thick conditions \citep{Leenaarts2013b}, the O\,{\sc i} and Si\,{\sc iv} lines are typically optically thin \citep{Lin2015}, and the C\,{\sc ii} line is generally thick but also with optically-thin conditions in some solar structures \citep{Rathore2015a,Rathore2015b}.

Numerical simulations of quiet Sun suggest that the Mg\,{\sc ii}\,h and k lines are excellent chromospheric diagnostics in the near UV, providing temperatures at various heights in the chromosphere \citep{Pereira2013}. For comparison, we study  both magnesium doublet lines in the present work, although their formation is very similar (with the k line forming a bit higher in the chromosphere due to a larger oscillator strength, by a factor of two, compared to the h line).
We refer the reader to \citet{Leenaarts2013a}, \citet{Leenaarts2013b}, \citet{Pereira2013}, and \citet{Pereira2015} for details on the formation of IRIS diagnostics.

Here, we aim at determining radiation temperatures (and intensities) of a few line-features of the IRIS spectra where each represents a particular formation height range in the solar chromosphere (and the low transition region). The line features, namely, the blue and red emission peaks and core of the magnesium k line (i.e. Mg\,{\sc ii}\,k2v, Mg\,{\sc ii}\,k2r, and Mg\,{\sc ii}\,k3, respectively) and of the magnesium h line (i.e. Mg\,{\sc ii}\,h2v, Mg\,{\sc ii}\,h2r, and Mg\,{\sc ii}\,h3) as well as the line cores of O\,{\sc i}~135.6~nm, C\,{\sc ii}~133.4~nm and Si\,{\sc iv}~139.4~nm, are computed with the MOSiC analysis tool \citep{Rezaei2017}. In short, after correcting the spectra for the orbital velocity of the satellite, among others, the various line features are determined at each pixel using a Gaussian fit, for which the degree of freedom gradually increases. It begins with a single Gaussian fit, then a higher level of multiple Gaussian (with a larger number of free parameters), if the previous one fails. Each fit provides an initial guess for the next one. The goodness of the fits at each step is evaluated by means of reduced chi-square statistics \citep{Bevington1992}. Only pixels with clear single and double peaked profiles are included in further analysis (i.e. complicated profiles are excluded). For the magnesium and carbon lines, the spectra have, in some places, only one emission peak (e.g. in the umbra), whereas in other areas they have two emission peaks with a central reversal. In the former case, the emission peak represents the line core, in the latter, the central reversal (i.e. k3 and h3 features). We note that no h2v, h2r, k2v, or h2r are identified in singly peaked profiles.

The IRIS raster images were spatially co-aligned with the ALMA scan by cross correlating the Mg\,{\sc ii}\,k2v and ALMA~1.3~mm images. In addition, we verified the alignments by means of a cross correlation of common features between the SDO/HMI continuum (which was separately co-aligned with ALMA) and the IRIS Mg\,{\sc ii} raster image at 283.2~nm (see Figure~\ref{fig:masks}b for the latter scan; also Section~\ref{subsec-sdoobs} for the co-alignment of ALMA and SDO images).

\subsection{Observing-time difference}
\label{subsec-obstime}

 Although the observing times of the entire FOV by ALMA and IRIS overlap, the two instruments scanned the FOV in different ways (ALMA did a mosaic of small FOVs with a zigzag-like path, whereas IRIS' slit scanned the entire FOV from east to west). The sampling time of each pixel was extracted for both ALMA and IRIS maps, thus, their observing-time difference could be determined. 

\begin{figure*}[!h]
\centering
    \includegraphics[width=.92\textwidth, trim = 0 0 0 0, clip]{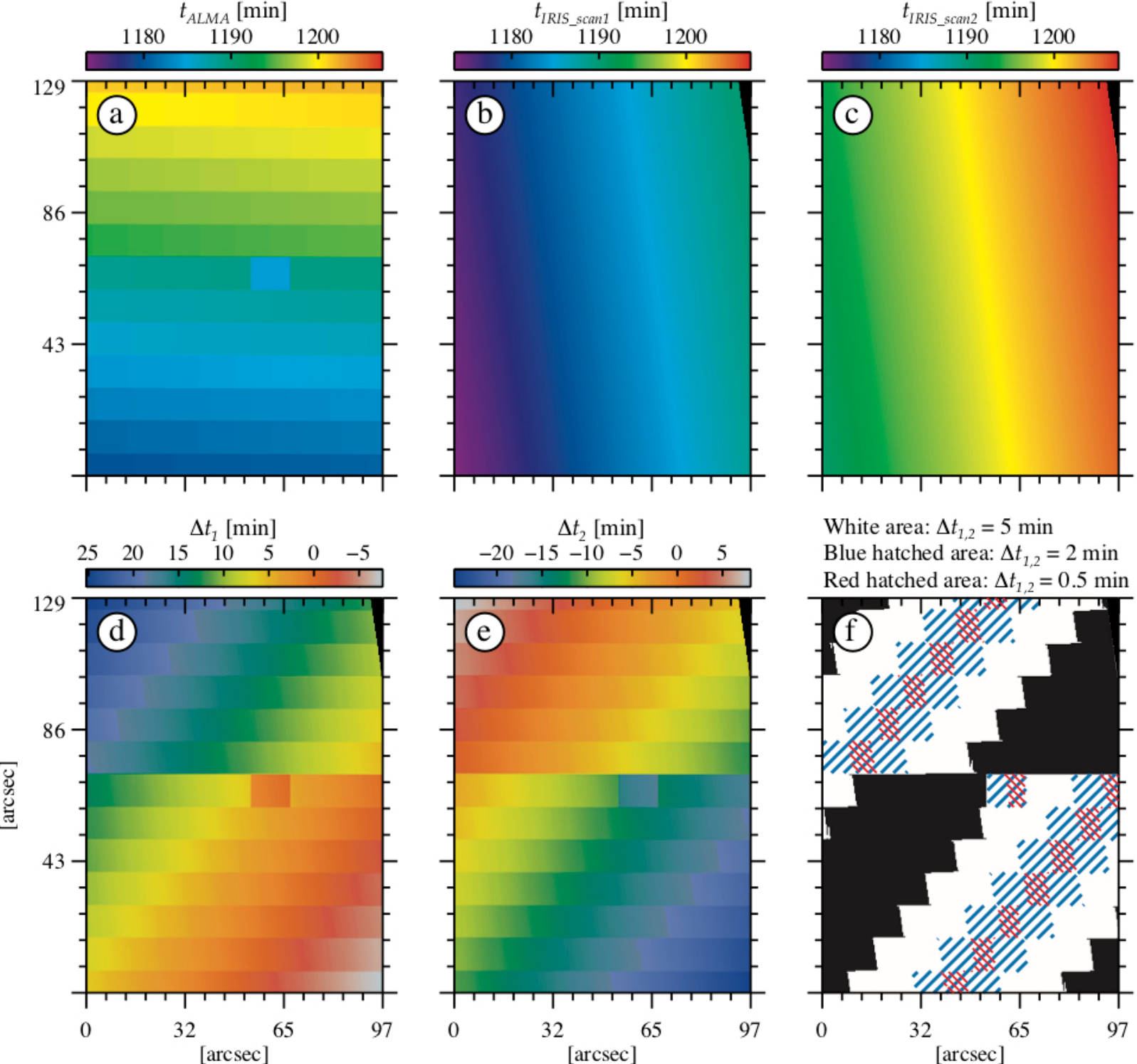}
  \caption{Sampling patterns of the ALMA image (a), the first (b) and the second (c) scans of IRIS, after spatial co-alignment. The colours in panels a-c indicate the observing times from the UT midnight.
  The observing-time difference maps between ALMA and the first and second IRIS scans ($\Delta t_{1}$ and $\Delta t_{2}$) are shown in panels (d) and (e), respectively). The positive values mean that the ALMA's sampling leads that of the IRIS. The colour in panels (d) and (e) represents the time-differences. Panel~(f) illustrates a mask where three regions with a maximum time-difference (indicated on top of the panel) are depicted. The mask is the result of combination of the first two IRIS scans (to cover the largest areas with small time-differences with ALMA observations).
  }
  \label{fig:obstime}
\end{figure*}

 The scanning patterns of the ALMA image as well as the first two scans of IRIS are illustrated in Figures~\ref{fig:obstime}a-c, respectively. The observing times (in minutes) from the UT midnight are colour coded. We note that the IRIS images were rotated with respect to the ALMA, as a result of the spatial alignment. The center of the original FOV of the ALMA was sampled first (but also later during the observations), before the zigzag pattern started from the lower left corner.
 
 Figures~\ref{fig:obstime}d and \ref{fig:obstime}e show the observing-time difference maps between the ALMA and the first and second scans of IRIS, respectively, where the IRIS' observing time (corresponding to the middle of each exposure at each pixel) is subtracted from that of the ALMA. The colours represent the observing-time differences. These figures reveal a relatively large time difference between sampling of a large fraction of the FOV by the two instruments. The highly dynamic Sun, particularly the solar chromosphere (of which most of the observations studied here are represented), varies in time-scales much shorter than 10~min. Hence, the time differences must be taken into account when one-to-one comparison between the two observations is performed. Luckily, it turned out that the two IRIS scans have sampled two different regions of the FOV closest in time with ALMA. Thus, a larger area with relatively small observing-time differences can be reconstructed when the two IRIS scans are combined accordingly. Such combined regions for various areas where the maximum time difference is 5~min (white area), 2~min (blue hatched area), and 0.5~min (red hatched area) are depicted in Figure~\ref{fig:obstime}f.
 
\subsection{SDO images}
\label{subsec-sdoobs}

Images from SDO/AIA \citep{Lemen2012} and SDO/HMI \citep{Schou2012}, corresponding to the ALMA scan, were also collected. The SDO--ALMA alignments were performed for the SDO images recorded close to the middle of the ALMA observations. We use a combination of the AIA 170~nm and 30.4~nm images to perform a precise spatial alignment with the ALMA scan. The combination of the two AIA channels (which sample the low chromosphere and the transition region, respectively) results in a similar scene to that observed with the ALMA's band 6, hence, facilitates the cross correlation of common features in the alignment procedure.

We also use the results of the Milne-Eddington Stokes inversion of the HMI data with the VFISV code \citep{Borrero2011} to obtain the photospheric magnetic field strength. 
This is used for partitioning the entire FOV into areas with different amoutns of (photospheric) magnetic fields, underlying their chromospheric (and transition region) counterparts sampled with ALMA and IRIS.

\subsection{Definition of regions}
\label{subsec-regions}

\begin{figure*}[!thp]
\centering
    \includegraphics[width=.92\textwidth, trim = 0 0 0 0, clip]{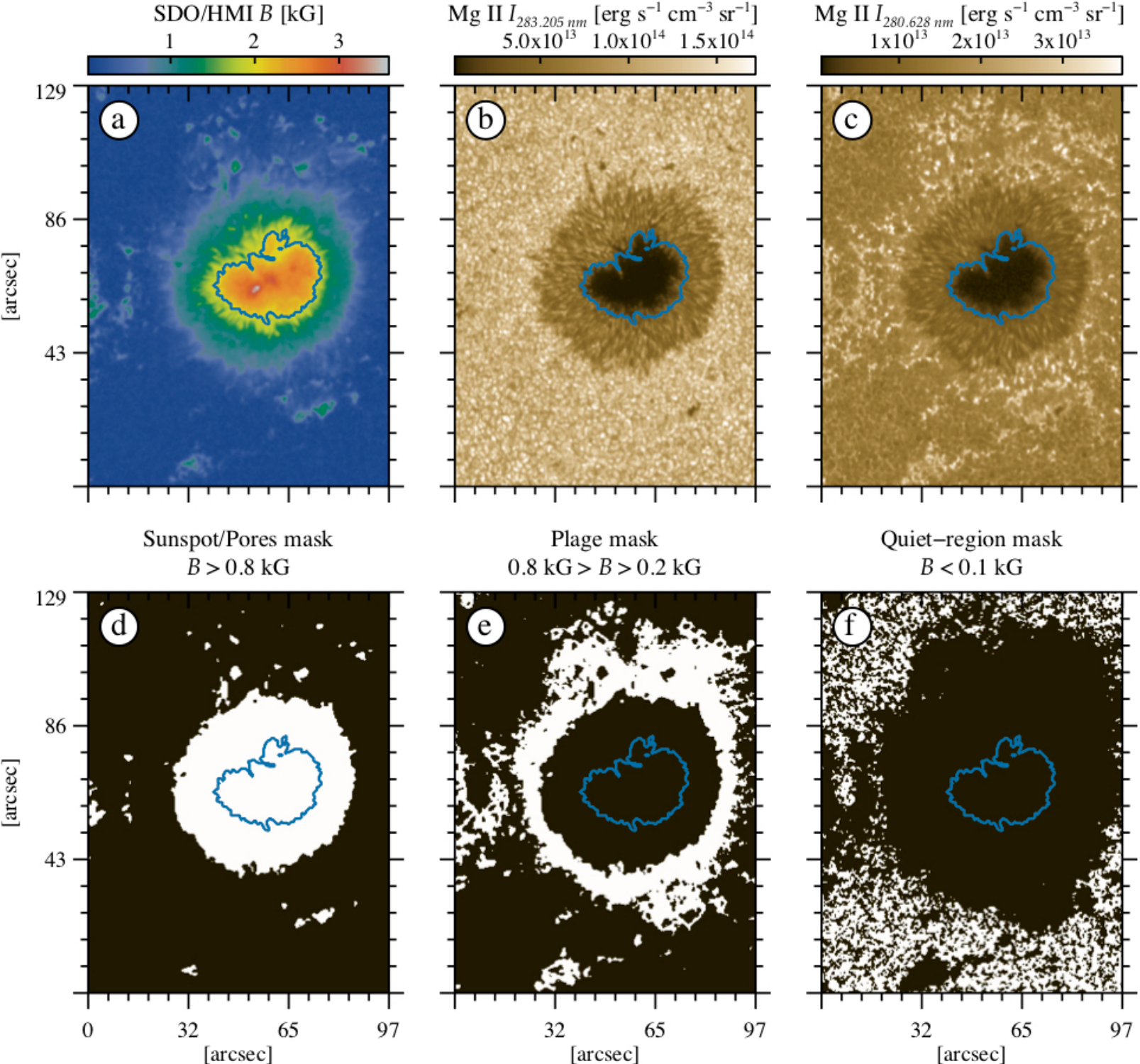}
  \caption{Regions of interest based on magnetic field strengths. Panel (a) shows the magnetic field map computed from SDO/HMI. Panels (b) and (c) illustrate two IRIS slit images representing the continuum and lower chromospheric heights, respectively. Three masks for the sunspot and pores, plage, and quasi-quiet areas are shown in panels (d--f). The white pixels of the masks are the areas of inclusion. The blue contour marks the umbra, including pixels with the field strengths larger than 2~kG.
  }
  \label{fig:masks}
\end{figure*}

The kinematics and brightness of different types of region in the  solar photosphere have been shown to be dependent on the magnetic flux \citep{Jafarzadeh2017,Borrero2017}, which is likely true for the chromospheric temperature, too.
The conventional approach to study such dependency is to partition the FOV into regions with distinct amounts of magnetic fields.
We define different regions based on the level of magnetic field strength, $B$, deduced from inversions of full Stokes SDO/HMI observations (of the magnetically sensitive line Fe\,{\sc i}~617.3~nm). A $B$-map, corresponding to the middle of the observing time window of the ALMA scan, is shown in Figure~\ref{fig:masks}a, ranging between 0.02-3.6~kG. Five areas are chosen: $B\geq2$~kG comprising only the umbra, $B\geq0.8$~kG including the whole sunspot and the small pores, 2~kG$>B\geq0.8$~kG that is similar to the previous region except the umbra and thus comprises the penumbra and pores, 0.8~kG$>B\geq0.2$~kG defining the plage areas, and $B<0.1$~kG to represent the quasi-quiet regions. Although the `umbra' and `penumbra and pores' are the two distinct (strong) magnetic regions of interest, for comparison with the less active areas, we have also included the ``sunspot and pores'' which includes the two former regions together. Also, we will show the radiation temperature distributions for the entire FOV, comprising all magnetic areas.

Figures~\ref{fig:masks}d--f show the sunspot and pore, plage, and quasi-quiet-Sun masks, respectively. The umbral region is marked by a contour on all maps in Figure~\ref{fig:masks}. For reference, two intensity images sampling the continuum and a height close to the temperature minimum or low chromosphere are also presented in Figures~\ref{fig:masks}b and \ref{fig:masks}c, respectively. The sunspot and small pores are clearly visible in panel b, whereas the plage and quasi-quiet regions are better seen in panel c.

\begin{figure*}[!thp]
\centering
    \includegraphics[width=.99\textwidth, trim = 0 0 0 0, clip]{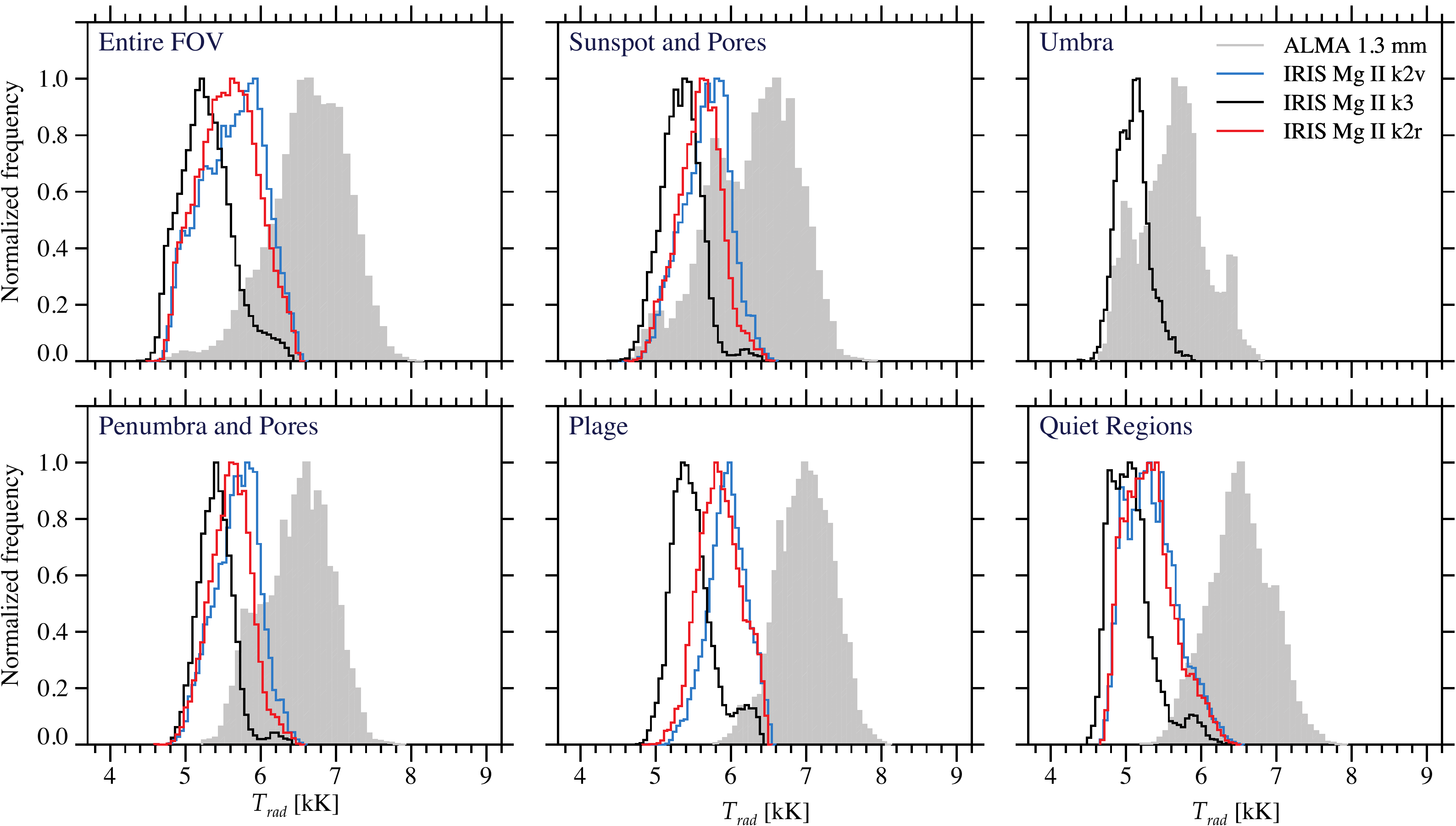}
  \caption{Distributions of radiation temperature of ALMA 1.3~mm (filled grey) and of various features of the IRIS Mg\,{\sc ii}\,k line at 279.6~nm (see the legend to the figure in the upper right panel). The histograms are plotted for the entire field-of-view and for the various isolated regions with different levels of magnetic fields (see main text).
  }
  \label{fig:histograms_alma_mgiik}
\end{figure*}

\begin{figure*}[!thp]
\centering
    \includegraphics[width=.99\textwidth, trim = 0 0 0 0, clip]{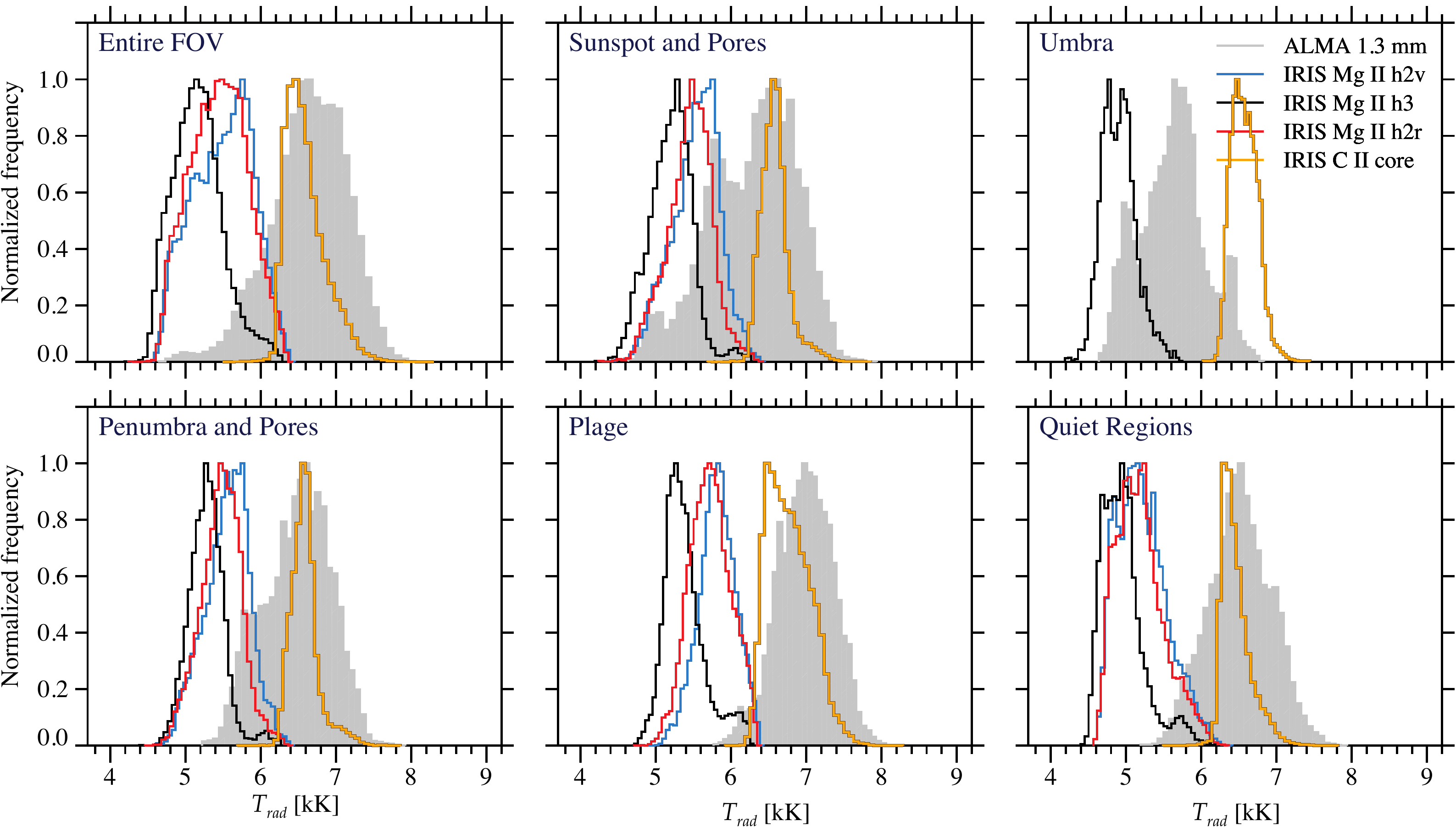}
  \caption{Same as Figure~\ref{fig:histograms_alma_mgiik}, but for the  Mg\,{\sc ii}\,h line as observed with IRIS at 280.4~nm as well as the IRIS C\,{\sc ii} 133.4~nm line-core.
  }
  \label{fig:histograms_alma_mgiih}
\end{figure*}

\begin{figure*}[!thp]
\centering
    \includegraphics[width=.99\textwidth, trim = 0 0 0 0, clip]{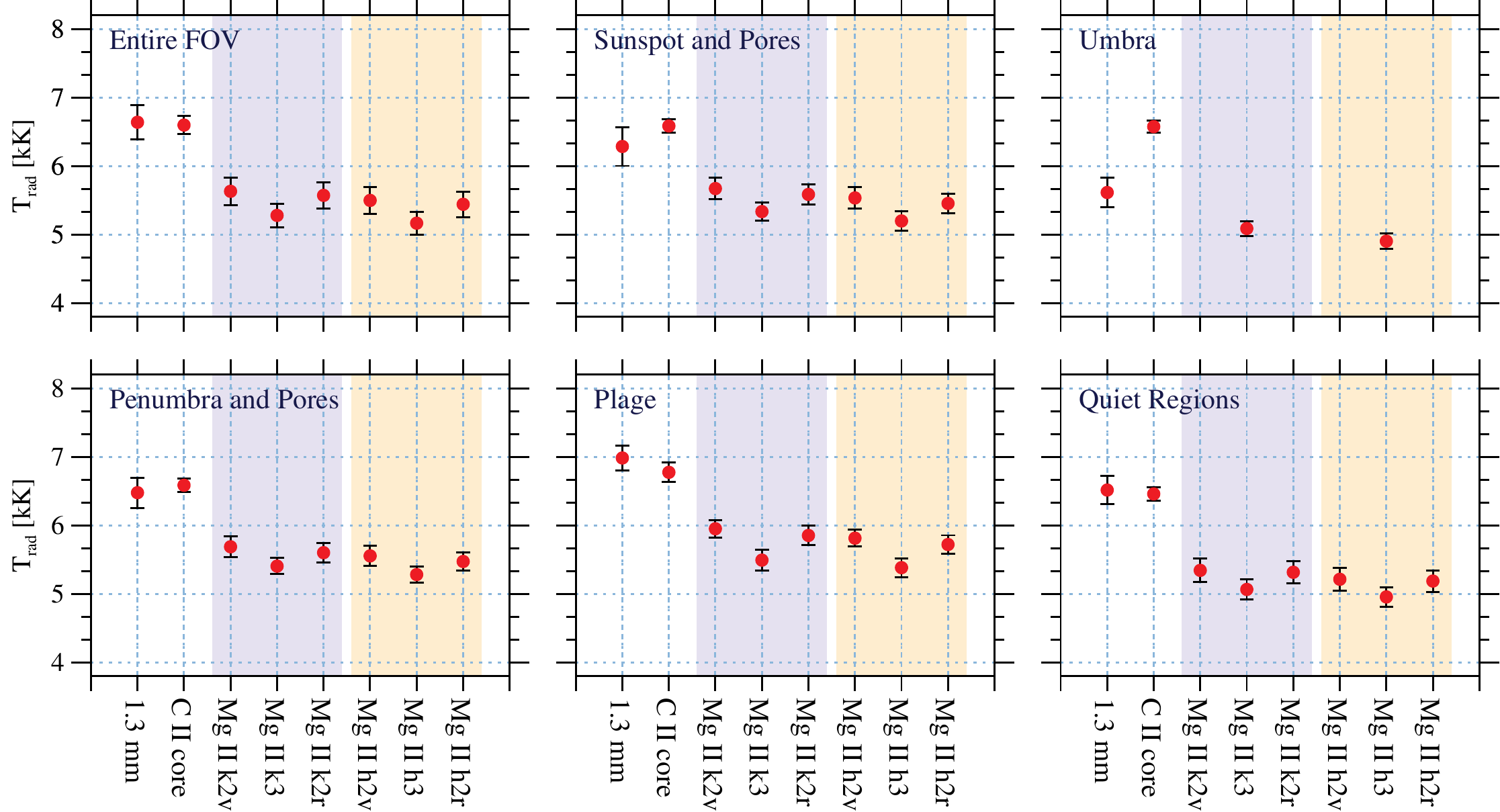}
  \caption{Comparison of the mean radiation temperature (red dots) retrieved from the millimetre and ultraviolet observations with ALMA and IRIS. The error bars indicate the standard deviations of the temperature distributions. For better visibility, the purple and yellow stripes have been depicted to mark the line features of the IRIS Mg\,{\sc ii}~k and Mg\,{\sc ii}~h, respectively.
  }
  \label{fig:histograms_comp}
\end{figure*}

\section{Analysis and results}
\label{sec-analys}

We study radiation temperatures throughout the solar chromosphere from observations at millimetre wavelengths with ALMA and in the UV with IRIS. The calibrated ALMA data product as output of the pipeline (both, the official ALMA pipeline and SoAP) is the brightness temperature (with an uncertainty in the order of 5\%). Thus, it directly enters our analysis with no further processing.

We calculate the near- and far-UV radiation temperature $T_{rad}$ in the solar chromosphere at a given wavelength $\lambda$ by solving the Planck function as
\begin{equation}
T_{rad}=\frac{h\, c/k\, \lambda }{\ln (((2\, h\, c^{2}/\lambda ^{5})/I)+1)}\,,
	\label{equ:planck}
\end{equation}
\noindent
where $h$, $c$, and $k$ are the Planck's constant, the speed of light in vacuum, and the Boltzmann’s constant, respectively. $I$ is the intensity, converted from the IRIS data numbers ($I_{DN}$;  in the spectral units):
\begin{equation}
I=\frac{I_\mathrm{DN}\, C_{f}\, (h\, c/\lambda )}{A_\mathrm{eff}\, \mathrm{Pix}_{y}\, \mathrm{Pix}_{\lambda }\, t_\mathrm{exp}\, W_\mathrm{slit}}\,,
	\label{equ:flux}
\end{equation}
\noindent
where $ C_{f}$ is the number of photons per data number, $A_\mathrm{eff}$ is the effective area, $\mathrm{Pix}_{y}$ and $\mathrm{Pix}_{\lambda}$ are respectively the sizes of the spatial and spectral pixels, $t_\mathrm{exp}$ is the exposure time, and $W_\mathrm{slit}$ is the slit width. These parameters were obtained using the latest version (version 4) of the IRIS radiometric calibration-routine (with an uncertainty of about 15\%). We note that, however, the radiation temperatures are estimated with an uncertainty of 2\% (due to the exponential of the Planck function).

The resulting radiation temperature maps for the particular line-parameters, computed using the MOSiC tool (see Section~\ref{subsec-irisobs}), are shown in Figures~\ref{fig:obs}b and in \ref{fig:obs}d--\ref{fig:obs}i. These include the line cores of C\,{\sc ii} 133.4~nm (b), the blue emission peak, line core, and red emission peak of Mg\,{\sc ii}\,k 279.6~nm (d--f) and of Mg\,{\sc ii}\,h 280.4~nm (g--i), respectively. 
For comparison, the intensity image of the IRIS Si\,{\sc iv} 139.4~nm is also presented in Figure~\ref{fig:obs}c. Blue pixels in the magnesium images are regions where the corresponding line features could not be determined (i.e. no clear single or double peaks were found; or no good fits were obtained). Only a few complicated profiles were found for the magnesium line-core images. The blue and red emission peaks could only rarely be identified in the umbra (the profiles are mostly singly peaked in the umbra). A few ``bad'' pixels in the C\,{\sc ii} and Si\,{\sc iv} line-core images were filled using a median filter. The O\,{\sc i} 135.6~nm line-core image was also determined using the MOSiC tool (not shown here). 
We note that, for clarity, the fiducial marks were removed (filled through the use of a median filter) from the slit images, prior to $T_{rad}$ calculations. We remind the reader that radiation temperature of the optically thin lines (i.e. O\,{\sc i} and Si\,{\sc iv}) are meaningless, as they are results of integration of the contribution function across a wide range of heights along the line of sight. Hence, their radiation temperatures are not calculated here.

A visual inspection of the $T_{rad}$ and intensity maps suggests the similarity of the ALMA 1.3~mm with the IRIS C\,{\sc ii} and Mg\,{\sc ii} lines, also in terms of fine structures. Comparisons between the $T_{rad}$ distributions and pixel-to-pixel relationships (of the radiation temperatures and intensities) are discussed in the following sections. In particular, we employ the masks defined in Section~\ref{subsec-regions} to also investigate dependency of the $T_{rad}$ distributions on the level of magnetic fields at the base of each region in the photosphere.

\subsection{Radiation temperature distributions}
\label{subsec-histograms}

Distributions of the radiation temperature measured at 1.3~mm by ALMA and at the multiple line features of the UV spectra by IRIS are plotted in Figures~\ref{fig:histograms_alma_mgiik} and \ref{fig:histograms_alma_mgiih}. In each figure, the ALMA's radiation temperature is compared with those from the UV channels, measured at all pixels corresponding to the entire FOV and the five different solar regions (sunspot and pores, umbra, penumbra and pores, plage, and quasi-quiet area). For simplicity and better comparison, the distributions of the Mg\,{\sc ii}\,k line features and of the Mg\,{\sc ii}\,h line features along with the C\,{\sc ii} line core are separately compared with those from ALMA in the two figures (i.e. Figures~\ref{fig:histograms_alma_mgiik} and \ref{fig:histograms_alma_mgiih}), respectively.
Please note that we aim here to present a statistical (not one-to-one) comparison of the radiation temperatures in similar magnetic regions. Hence, the IRIS radiation temperatures correspond to those computed from the second scan of the IRIS raster with its original spatial resolution. This means that there can be a substantial time difference between the observations in part of the FOV. 
The one-to-one comparison of the radiation temperatures are provided in Section~\ref{subsec-scatterplots} between the ALMA map and the IRIS images degraded to that of the ALMA, and only for pixels with small observing-time differences.

The ranges, means, and standard deviations of the radiation temperature distributions of the different lines, and for the various areas, are summarised in Table~\ref{table:stat_obs}. The mean radiation temperatures of the millimetre and of the seven UV channels are also compared in Figure~\ref{fig:histograms_comp} for the six different solar regions under study. The error bars in Figure~\ref{fig:histograms_comp} represent the standard deviations of the $T_{rad}$ histograms. For clarity, the two Mg\,{\sc ii} lines are marked with two coloured stripes.
We note that no statistics are shown for the blue and red emission peaks of the Mg\,{\sc ii} lines in the umbra, since almost no double emission peaks were identified there.

The brightness temperature at 1.3~mm varies between 4.6--8.1~kK in the entire FOV with an average and standard variation of 6.6~kK and 0.5~kK, respectively. Its mean value is the smallest in the umbra ($5.6\pm0.4$~kK; where the uncertainty is the standard deviation of the distribution), and is the largest in the plage region ($7.0\pm0.4$~kK). 
In comparison with the UV radiation temperatures, the ALMA distribution has the best overlap with that from the C\,{\sc ii} in all regions except in the umbra where the far-UV line poses higher $T_{rad}$ compared to that from the mm band. However, we note that the radiation temperature distributions of ALMA 1.3~mm are always wider (by a factor of about two larger standard deviations; the exception is in the plage area) than those of C\,{\sc ii}. This may suggest that the ALMA 1.3~mm sub-band includes contributions from a wider range of chromospheric heights than the C\,{\sc ii} line.

\begin{table*}[!thp]
\begin{center}
\caption{Summary of distributions of radiation temperature at millimetre and ultraviolet wavelengths from ALMA and IRIS.}
\vspace{-2.5mm}
\label{table:stat_obs}
\setlength{\tabcolsep}{0.77em}   
\renewcommand{\arraystretch}{1.3}         
\begin{tabular}{l | l c c c c c c c c}
\Xhline{0.8pt}
\\[-1.25em]
\Xhline{0.8pt}
\\[-1.1em]
Region~\tablefootmark{$a$} & Parameter & ALMA & C\,{\sc ii} & Mg\,{\sc ii}\,k2v & Mg\,{\sc ii}\,k3 & Mg\,{\sc ii}\,k2r & Mg\,{\sc ii}\,h2v & Mg\,{\sc ii}\,h3 & Mg\,{\sc ii}\,h2r \\
 & [kK] & 1.3~mm & 133.4~nm &  &  &  &  &  & \\
\Xhline{0.8pt}
\multirow[t]{3}{*}{Entire} & mean & 6.6 & 6.6 & 5.6 & 5.3 & 5.6 & 5.5 & 5.2 & 5.4 \\
{FOV} & $\sigma$~\tablefootmark{$b$} & 0.5 & 0.3 & 0.4 & 0.3 & 0.4 & 0.4 & 0.3 & 0.4 \\
{} & range & 4.6--8.1 & 5.5--8.2 & 4.5--6.6 & 4.4--6.4 & 4.5--6.5 & 4.3--6.4 & 4.2--6.3 & 4.2--6.4 \\
\Xhline{0.8pt}
\multirow[t]{3}{*}{Sunspot} & mean & 6.3 & 6.6 & 5.7 & 5.3 & 5.6 & 5.5 & 5.2 & 5.5 \\
{\& Pores} & $\sigma$ & 0.6 & 0.2 & 0.3 & 0.3 & 0.3 & 0.3 & 0.3 & 0.3 \\
{} & range & 4.6--7.8 & 5.7--7.8 & 4.5--6.6 & 4.4--6.4 & 4.6--6.5 & 4.3--6.4 & 4.2--6.3 & 4.2--6.4 \\
\Xhline{0.8pt}
\multirow[t]{3}{*}{Umbra} & mean & 5.6 & 6.6 & -- & 5.1 & -- & -- & 4.9 & -- \\
{} & $\sigma$ & 0.4 & 0.2 & -- & 0.2 & -- & -- & 0.2 & -- \\
{} & range & 4.6--6.8 & 6.0--7.4 & -- & 4.4--5.9 & -- & -- & 4.2--5.7 & -- \\
\Xhline{0.8pt}
\multirow[t]{3}{*}{Penumbra} & mean & 6.5 & 6.6 & 5.7 & 5.4 & 5.6 & 5.6 & 5.3 & 5.5 \\
{\& Pores} & $\sigma$ & 0.4 & 0.2 & 0.3 & 0.2 & 0.3 & 0.3 & 0.2 & 0.3 \\
{} & range & 5.2--7.8 & 5.7--7.8 & 4.8--6.6 & 4.4--6.4 & 4.6--6.5 & 4.6--6.4 & 4.4--6.3 & 4.5--6.4 \\
\Xhline{0.8pt}
\multirow[t]{3}{*}{Plage} & mean & 7.0 & 6.8 & 6.0 & 5.5 & 5.9 & 5.8 & 5.4 & 5.7 \\
{} & $\sigma$ & 0.4 & 0.3 & 0.3 & 0.3 & 0.3 & 0.2 & 0.3 & 0.3 \\
{} & range & 5.8--8.1 & 5.9--8.2 & 4.8--6.5 & 4.7--6.4 & 4.8--6.5 & 4.7--6.4 & 4.7--6.3 & 4.7--6.3 \\
\Xhline{0.8pt}
\multirow[t]{3}{*}{Quasi-quiet} & mean & 6.5 & 6.5 & 5.3 & 5.1 & 5.3 & 5.2 & 5.0 & 5.2 \\
{Regions} & $\sigma$ & 0.4 & 0.2 & 0.3 & 0.3 & 0.3 & 0.3 & 0.3 & 0.3 \\
{} & range & 5.2--7.9 & 5.5--7.8 & 4.6--6.5 & 4.5--6.4 & 4.7--6.5 & 4.6--6.4 & 4.4--6.2 & 4.6--6.3 \\
\Xhline{0.8pt}
\end{tabular}
\end{center}
\hspace{1mm}
\vspace{-5.0mm}
\tablefoot{
\tablefoottext{$a$}{Solar regions defined based on strength of the underlying magnetic fields (see Section~\ref{subsec-regions} and Figure~\ref{fig:masks}).}
\tablefoottext{$b$}{Standard deviations of the distributions.}
}
\end{table*}

The emission peaks and line cores of the two magnesium lines pose the smallest (similar) mean radiation temperatures in our study, the line cores have slightly lower $T_{rad}$ compared to the emission peaks. This is in agreement with an IRIS diagnostic study (from a numerical simulations with the Bifrost code) where the source function was found to be typically decoupled from the Plank function at the Mg\,{\sc ii} line cores \citep{Leenaarts2013b}. 
Among all the distinct areas under study, the Mg\,{\sc ii} lines have the largest mean $T_{rad}$ in the plage region and the smallest in the umbra (i.e. a similar behaviour to those of the ALMA 1.3~mm). The $T_{rad}$ distributions of the Mg\,{\sc ii} lines have small overlaps with that of the ALMA Band 6.

We remind the reader that the $T_{rad}$ of the Mg\,{\sc ii} lines (and of the C\,{\sc ii} line) is not equal to their actual gas temperature \citep{Leenaarts2013b,Rathore2015a}, thus the comparison of their $T_{rad}$ with ALMA should be done with great caution (ALMA's $T_{rad}$ is a good representative of the gas temperature at the heights sampled by the mm line; the non-LTE formation of the UV lines versus the LTE formation of the ALMA Band 6). Furthermore, different heights of formation of the various spectral lines can play a role in the differences observed between the $T_{rad}$ distributions. When comparing the histograms presented in this section, one should also keep in mind that there can be a substantial time differences between observations at UV and mm wavelengths in part of the FOV.

\subsection{Relationships between ALMA and IRIS}
\label{subsec-scatterplots}

Next, we compare the brightness temperature of the ALMA 1.3~mm with those computed from the IRIS UV channels by means of scatter plots. To this end, we plot pixel-to-pixel relationships of the radiation temperature of the seven UV channels discussed in the previous section, as well as the line-core intensities of the IRIS Si\,{\sc iv} and O\,{\sc i}, versus the ALMA Band~6 brightness temperature.
Prior to the comparison, the IRIS images are (1) convolved with the PSF of the ALMA map (i.e. equivalent to a $2\farcs2\times0\farcs7$ Gaussian kernel), and (2) re-sampled to the same pixel size as in the ALMA map. Hence, all maps for the comparison have the same spatial resolution (i.e. the spatial resolution of the ALMA Band~6 image). More importantly, we only include pixels, from either of the first two IRIS scans, which have a maximum observing-time difference of two minutes with their corresponding pixels in the ALMA map (i.e. areas marked with blue hatched in Figure~\ref{fig:obstime}f; see Section~\ref{subsec-obstime} about the time differences). Although this selection limits us to only 28\% of the entire FOV, the one-to-one comparison is free from temperature changes happening on time scales longer than 2~min. We note that large temperature variations of, for instance, up to $\approx$1kK, occurring within 2-3 minutes, have been observed in the upper chromosphere of sunspots \citep{delaCruz2013b,Joshi2018}. Thus, a large time-difference (e.g. larger than 2 min) between an observed pixel in the two images can significantly bias the one-to-one comparison. Furthermore, we also investigate the relationships for pixels with a more conservative time difference of maximum 0.5~min between sampling of the UV diagnostics and of the millimetre channel (i.e. areas marked with red hashed in Figure~\ref{fig:obstime}f; about 8\% of the entire FOV).

\begin{figure*}[!thp]
\centering
    \includegraphics[width=.99\textwidth, trim = 0 0 0 0, clip]{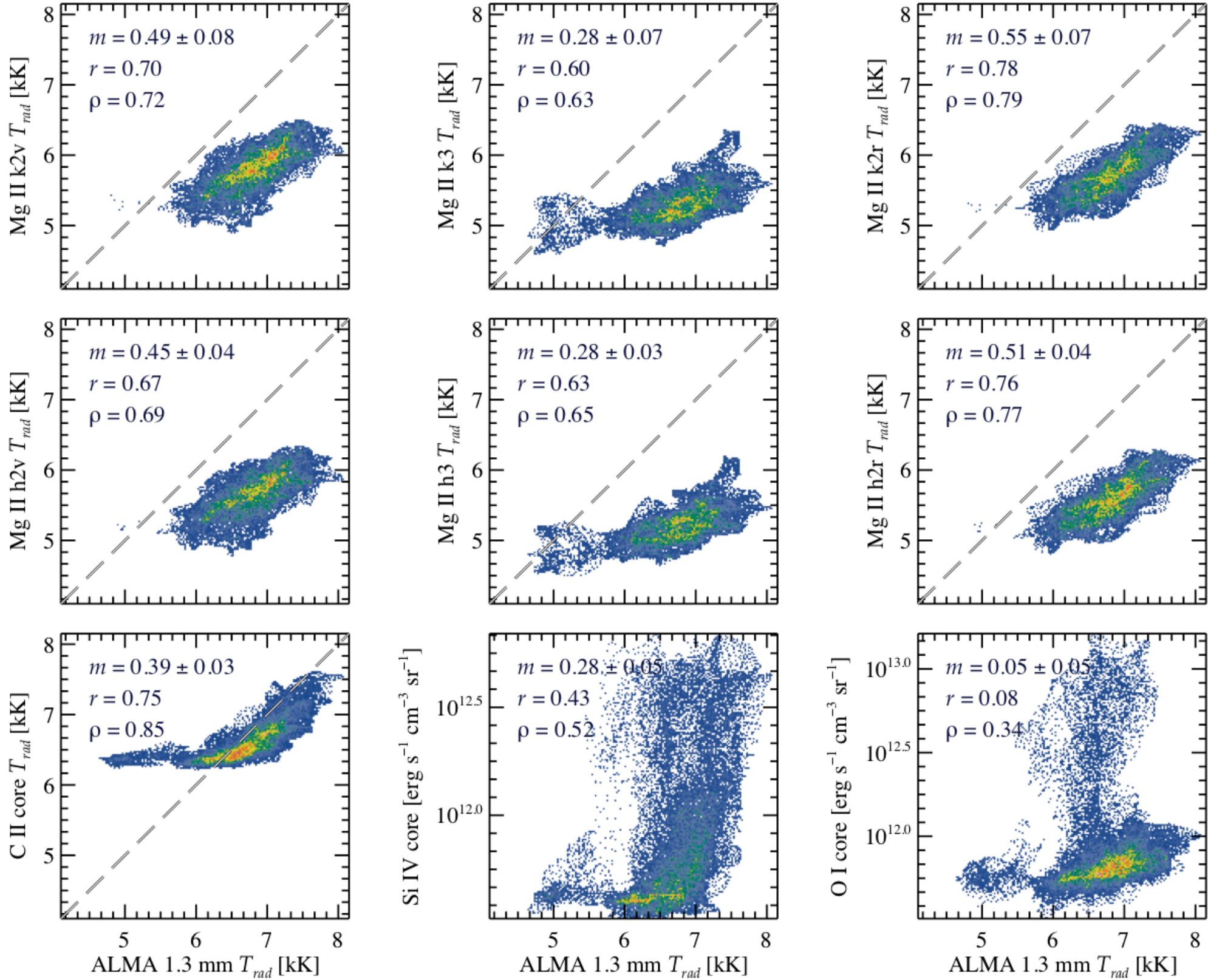}
  \caption{Scatter plots of radiation temperatures derived from various near and far ultraviolet line-features from IRIS versus those obviated with ALMA at 1.3~mm. The plots represent the pixel-to-pixel relationships for the pixel where their samling time difference are equal to or smaller than two minutes, corresponding to the blue hatched area in Figure~\ref{fig:obstime}f. The colours represent density of the scatter points (with the highest density number being coloured red and the lowest density dark blue). At each plot, the regression slope of the linear fit ($m$), the Pearson correlation coefficient ($r$), and the Spearman's rank correlation coefficient ($\rho$) are also indicated.
  }
  \label{fig:scatterplots_all}
\end{figure*}

\begin{figure*}[!thp]
\centering
    \includegraphics[width=.99\textwidth, trim = 0 0 0 0, clip]{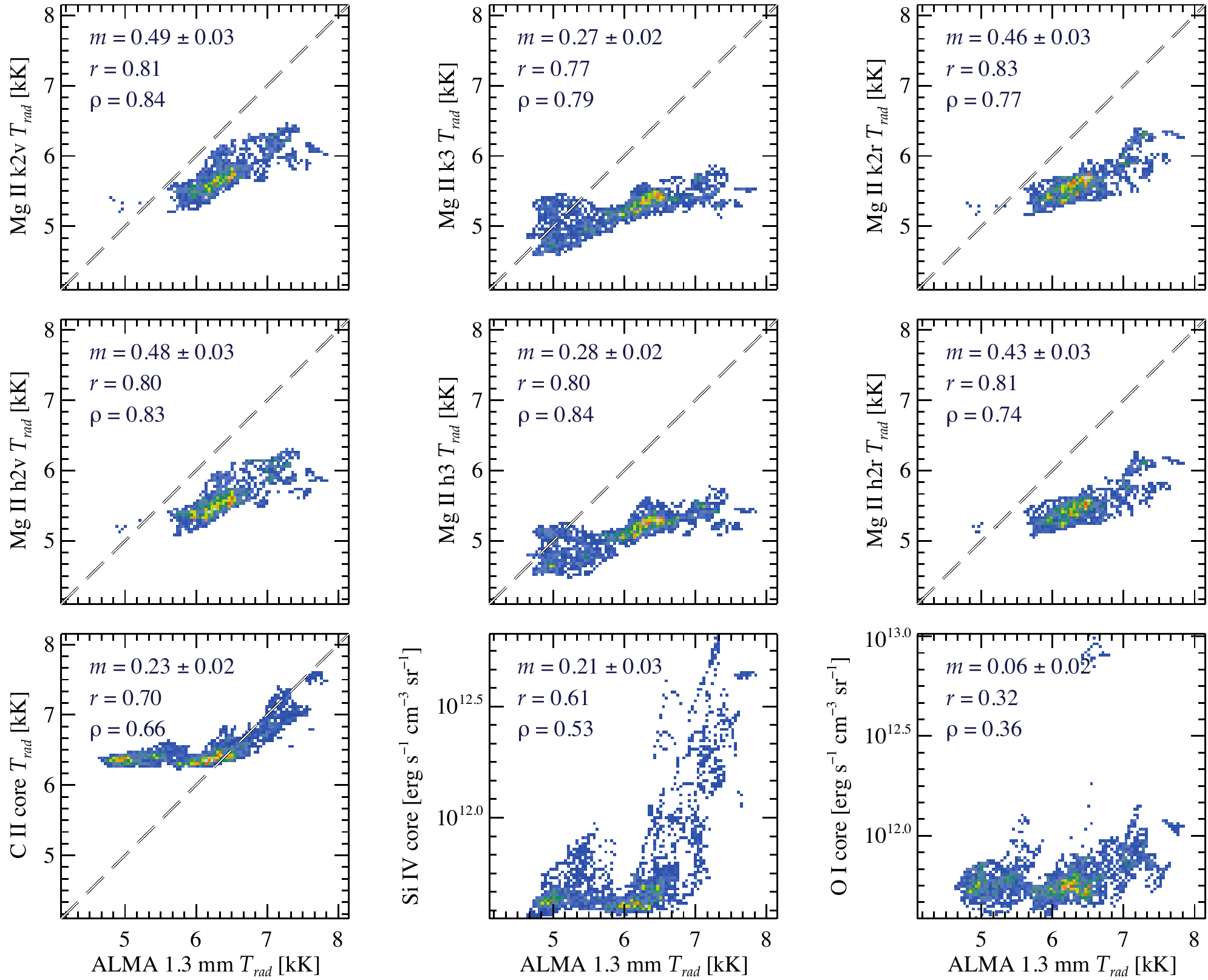}
  \caption{Same as Figure \ref{fig:scatterplots_all}, but limited to pixels included in the sunspot and the small pores.
  }
  \label{fig:scatterplots_spotpore}
\end{figure*}

\begin{figure*}[!thp]
\centering
    \includegraphics[width=.99\textwidth, trim = 0 0 0 0, clip]{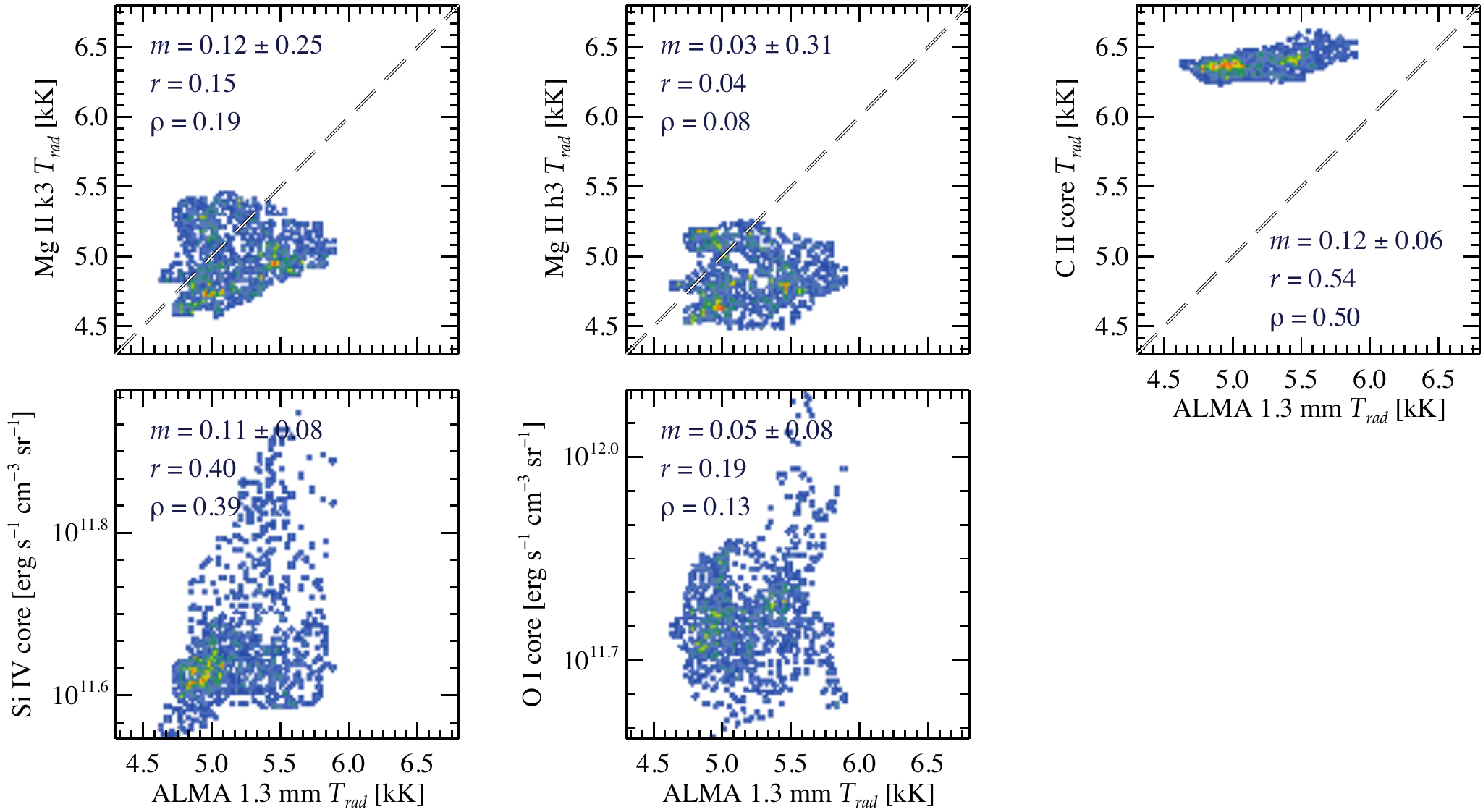}
  \caption{Same as Figure \ref{fig:scatterplots_all}, but limited to pixels in the umbral region.
  }
  \label{fig:scatterplots_umbra}
\end{figure*}

\begin{figure*}[!thp]
\centering
    \includegraphics[width=.99\textwidth, trim = 0 0 0 0, clip]{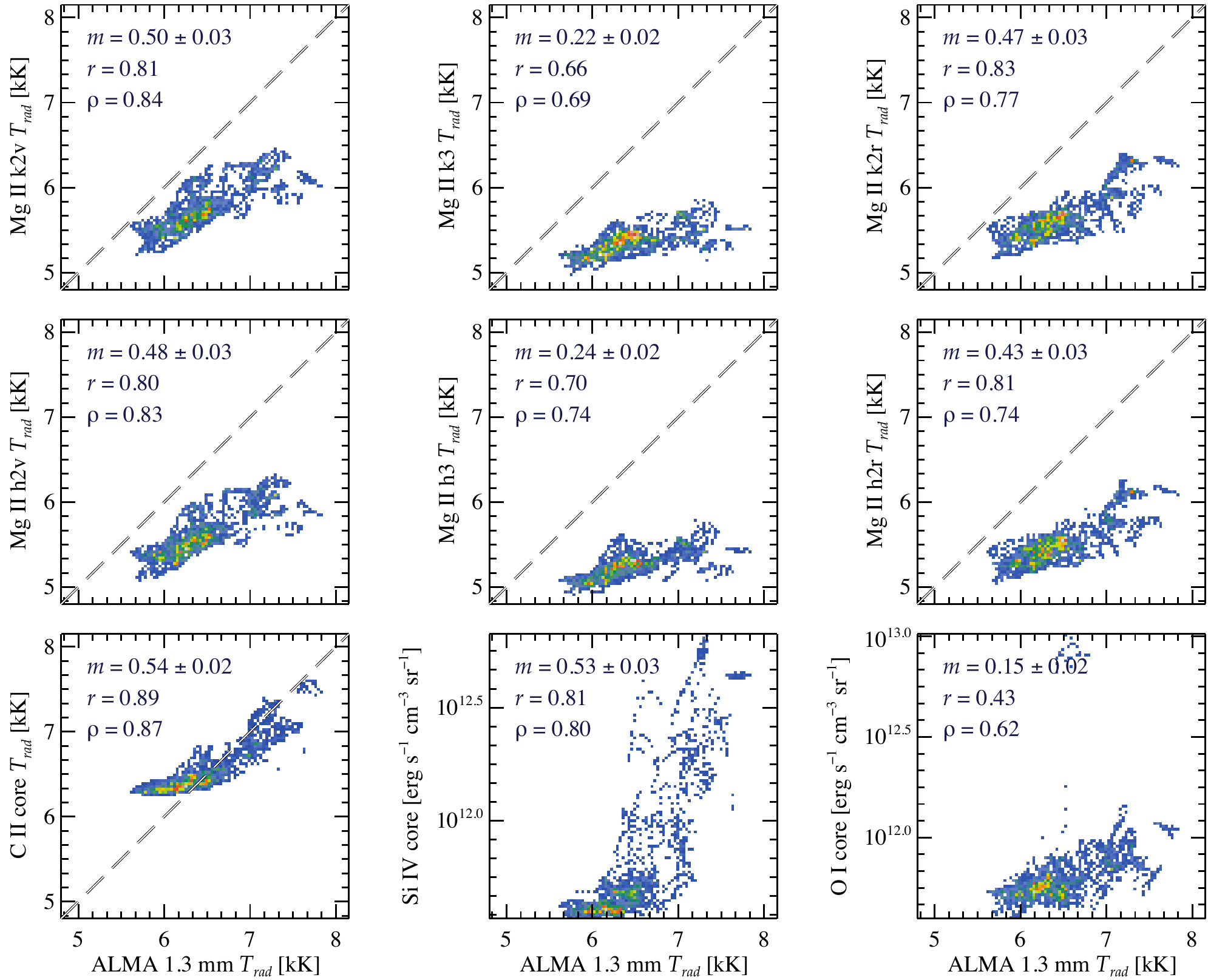}
  \caption{Same as Figure \ref{fig:scatterplots_all}, but for the penumbra and the small pores.
  }
  \label{fig:scatterplots_penumbrapore}
\end{figure*}

\begin{figure*}[!thp]
\centering
    \includegraphics[width=.99\textwidth, trim = 0 0 0 0, clip]{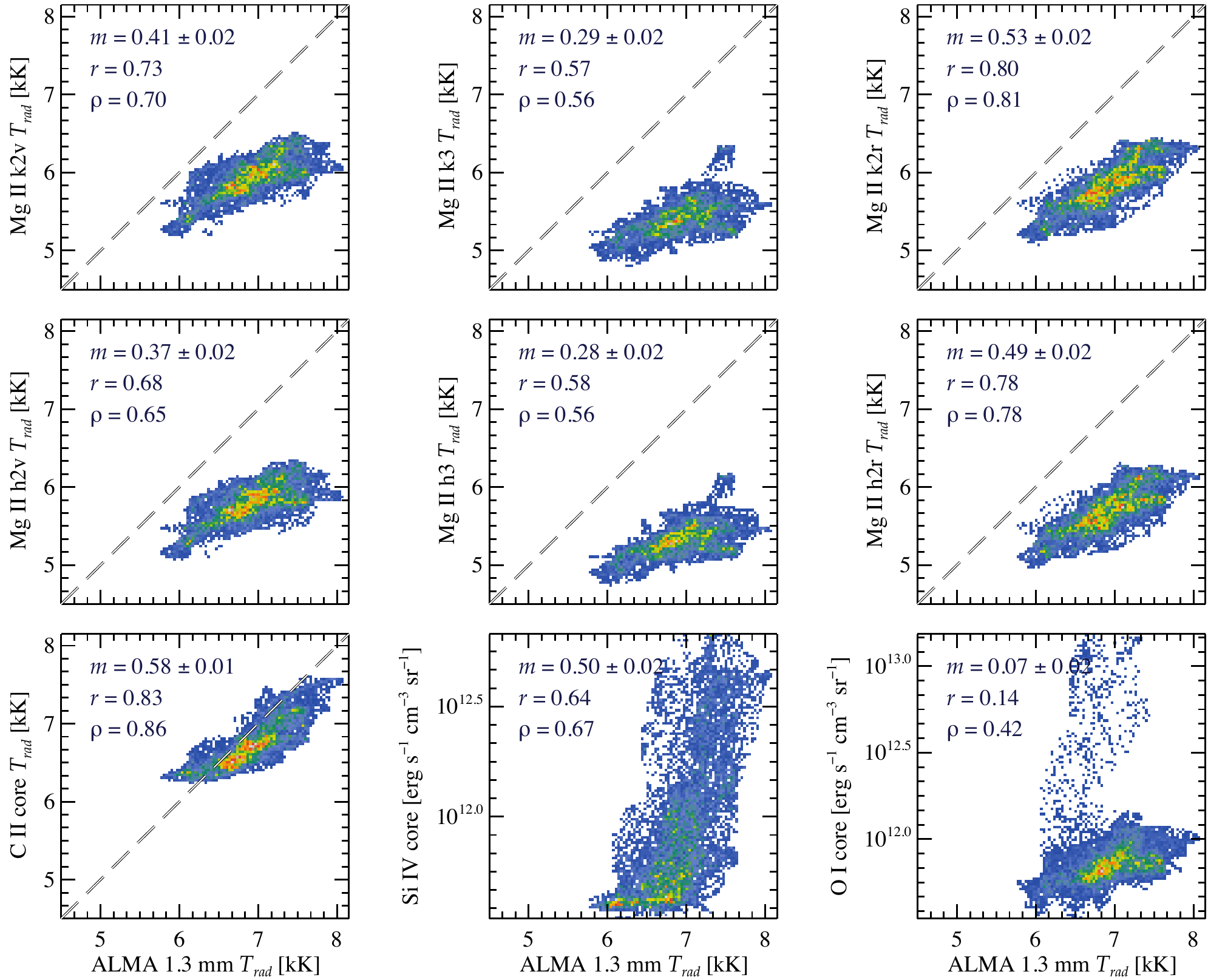}
  \caption{Same as Figure \ref{fig:scatterplots_all}, but limited to pixels of the plage areas.
  }
  \label{fig:scatterplots_plage}
\end{figure*}

\begin{figure*}[!thp]
\centering
    \includegraphics[width=.99\textwidth, trim = 0 0 0 0, clip]{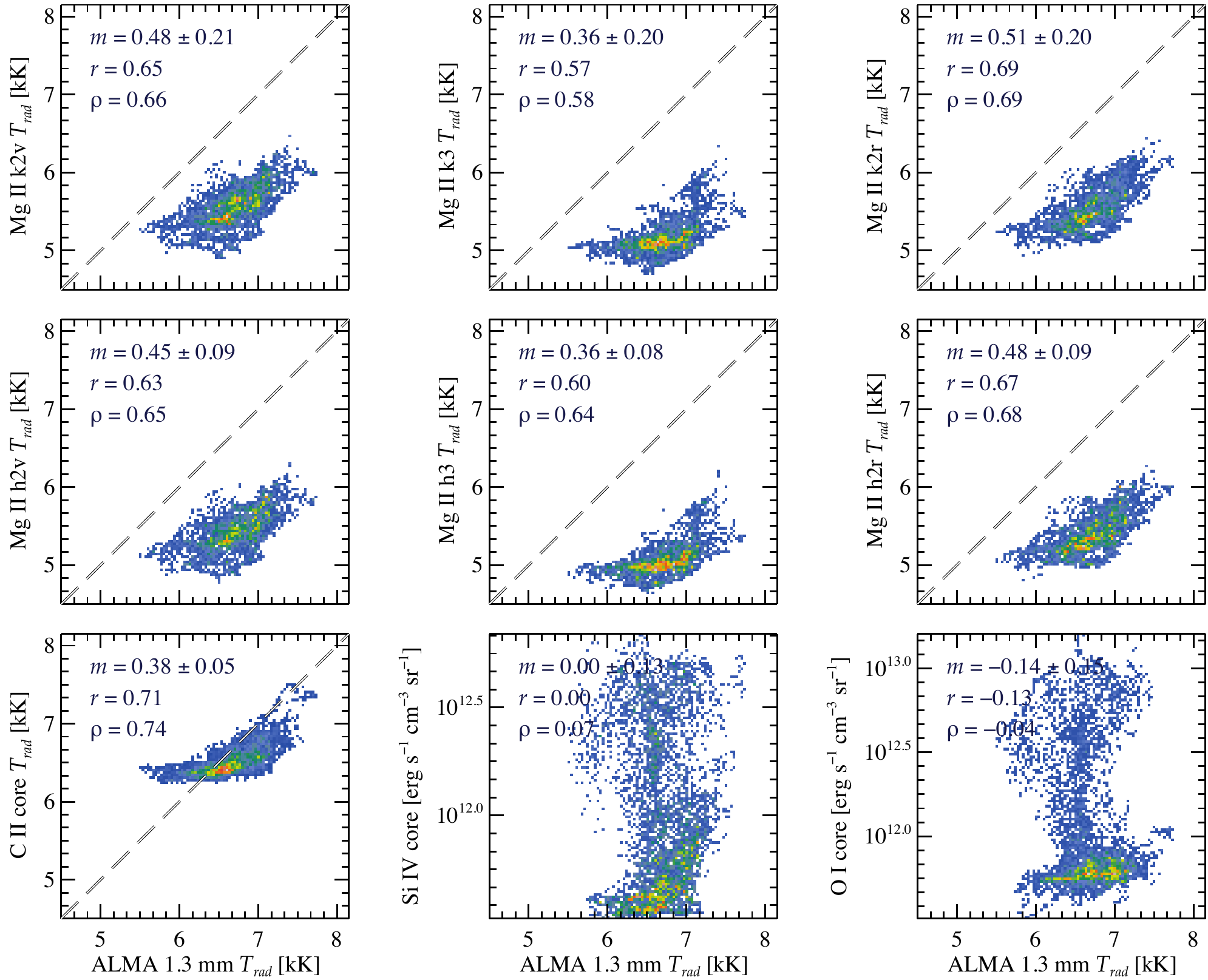}
  \caption{Same as Figure \ref{fig:scatterplots_all}, but for the quasi-quiet regions.
  }
  \label{fig:scatterplots_qs}
\end{figure*}

Figures~\ref{fig:scatterplots_all}-\ref{fig:scatterplots_qs} show these relationships for the six different regions introduced in Section~\ref{subsec-regions} (for which, their temperature distributions were discussed in Section~\ref{subsec-histograms}). The colour of the scatter points represents the density number of the scatter plots (from red and yellow to green and dark-blue indicating the highest concentrations to the lowest). 
We note that no scatter plots are shown for the blue and red emission peaks of the magnesium lines in the umbra (Figure~\ref{fig:scatterplots_umbra}), since the double emission peaks were only identified in a few pixels there.

In order to test for a potential linear relationship between the radiation temperature (or intensity) values of the UV diagnostics and that of the ALMA, the Pearson correlation coefficient $r$ is calculated for each pair. The coefficient $r$ varies between $\pm1$, where $r=+1$ represents a perfect linear correlation and $r=-1$ a perfect linear anti-correlation. 
Furthermore, the linear regression coefficient of the linear fit $m$, which measures the steepness of a linear relationship, is calculated.
However, in several cases, the relationships between radiation temperatures of the pair lines appear to be nonlinear. Hence, we also calculate the Spearman's rank correlation coefficient $\rho$, which better models non-linear relationships and is not sensitive to outliers.
The values of $m$, $r$, and $\rho$ are printed in all scatter plots in Figures~\ref{fig:scatterplots_all}--\ref{fig:scatterplots_qs}. 
The dashed lines in all $T_{rad}$-$T_{rad}$ plots indicate the $y=x$ line. It is evident that even with a relatively large correlation coefficients (and/or large slopes), there is a large offset between the scatter clouds and the $y=x$ line, suggesting a disagreement between each of the two distributions under comparison. The only exception is when the temperatures of the IRIS C\,{\sc ii} line and ALMA 1.3~mm are compared, despite their non-linear relationships. However, it is important to note that the disagreements are between the radiation temperatures. It has been shown that the actual gas temperatures measured by the Mg\,{\sc ii} and C\,{\sc ii} diagnostics are in fact higher than their $T_{rad}$, resulting in an offset between their linear trend of $T_{gas}$-$T_{rad}$ scatter points and the $y=x$ line (the source and Planck functions are, to some extent, decoupled in those cases; \citealt{Leenaarts2013b,Rathore2015a}).

We note that if the time difference had not be considered, different correlation coefficients and/or slopes would be found.

The majority of the $T_{rad}$-$T_{rad}$ scatter plots (for all solar regions, except the umbra) show a moderate-to-good positive linear correlation (i.e. $0.5 < r < 0.9$). This suggests that the relative low (or high) temperatures in the ALMA 1.3~mm map are associated with correspondingly low (or high) temperatures in the IRIS near and far UV channels. This is not surprising, since most of the spectral diagnostics discussed here are supposedly formed in the mid-to-high chromosphere (or the low transition region) and should thus probe similar temperature distributions. However, there may be height differences between sampling of the different diagnostic in various magnetic regions.
The scatter of the data points could imply that the various lines have different contribution functions to the radiation, meaning that the measured intensities are built up differently along the line of sight. 
Different diagnostics are expected to probe the thermal structure of the chromosphere in a slightly different way over effectively slightly different height ranges, where some are affected by deviations from equilibrium conditions (e.g. non-LTE) more than others. Finally, uncertainties in the measurements and the derivation of temperatures from them contribute to the scatter. 
In particular, the scatter in the Si\,{\sc iv} and O\,{\sc i} plots could partially be due to the optically-thin formation of these lines as emergent intensity is integrated over a large region along the line of sight.

By inspecting the 3D radiation MHD simulations from the Bifrost code \citep{Gudiksen2011,Carlsson2016}, we found a similar behaviour in the solar chromosphere (in quiet and enhanced network regions). The larger height difference between two distinct layers in the model chromosphere resulted in a wider scatter in their pixel-to-pixel relationships of gas temperatures (while the positive correlations always maintained). 

We note that we cannot completely rule out non-linear relationships between the radiation temperatures (or intensities) derived from the various UV bands (particularly, from the far-UV channels) and that from ALMA band 6 in individual pixels. However, linear relationships in an average sense are seen in the observations (see Figures.~\ref{fig:scatterplots_all}--\ref{fig:scatterplots_qs}) and thus imply an average behaviour in the mid-to-upper solar chromosphere, which is mostly independent of the level of magnetic fields. Moreover, the Spearman correlation coefficients (which better models non-linear relationships) were also provided for all plots.

\begin{figure*}[!thp]
\centering
    \includegraphics[width=.99\textwidth, trim = 0 0 0 0, clip]{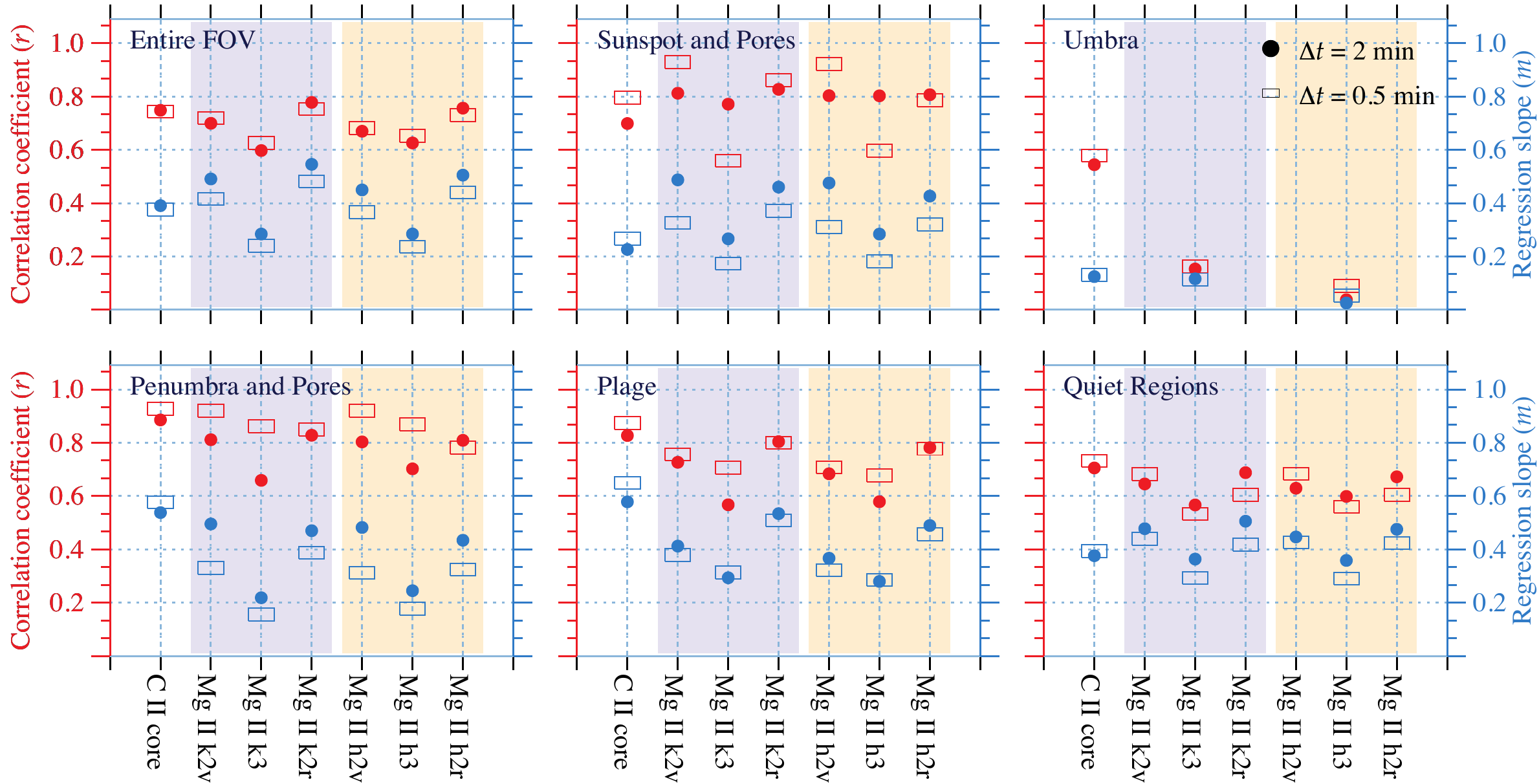}
  \caption{Comparison of Pearson correlation coefficients $r$ (red; left $y$-axis) as well as regression slopes of linear fits $m$ (blue; right $y$-axis) in different solar regions between radiation temperatures of various UV pixels and their corresponding ALMA~1.3~mm. The filled circles and open rectangles indicate the parameters for two sets of pixels with maximum observing time differences of 2 and 0.5~min, respectively.
  }
  \label{fig:correlations_comp}
\end{figure*}

Furthermore, the rather wide spread of the points in (almost) all the plots as well as the offsets between their linear trends and the $y=x$ line could still partly be due to the sampling time-differences remained in the above analyses (i.e. the maximum 2~min time differences between observations of the UV and mm diagnostics). To investigate the effect of rapid solar evolution in time scales shorter than 2~min, we also repeated the analyses with a maximum time difference of 0.5~min. However, this limits the number of data points, which in turn may introduce some uncertainties in determining the correlations. The correlation coefficients $r$ as well as slopes of the scatter plots $m$ for pixels with a maximum time difference of 0.5~min have somewhat small differences compared to those with larger time lags of up to 2~min. Figure~\ref{fig:correlations_comp} visualizes the comparison of $r$ (red) and $m$ (blue) parameters between the two analyses, where the filled circles corresponds to the $\Delta t=2$~min comparison (i.e. those shown in Figures~\ref{fig:scatterplots_all}--\ref{fig:scatterplots_qs}) and the open rectangles represents those for $\Delta t=0.5$~min. It is evident that reducing the observing time differences has resulted in a larger Pearson correlations coefficients in most cases, but smaller slopes. The smaller time difference has, however, insignificant effect on the offsets between the linear trends and the $y=x$ line in the scatter plots. The offsets, as previously noted, are likely due to the non-LTE formation of the UV diagnostics, thus their $T_{rad}\neq T_{gas}$, and/or differences in their heights of formation. We note that the distributions of the radiation temperature are generally similar in both cases of $\Delta t=2$ and 0.5~min, but the scatter points are less spread in the latter case. This, additionally, stresses the importance of accounting for time differences for one-to-one comparisons.

\subsubsection{Agreements between temperature distributions}
\label{subsec-kstest}

To investigate how the distributions of the radiation temperatures, compared in the scatter plots in Section~\ref{subsec-scatterplots} (i.e. from the similar spatial-resolution data; sampled within two minutes in time), are compared, we also employ the Kolmogorov–Smirnov (K-S) test. In this statistical approach, the maximum differences ($d$) between cumulative distribution functions (CDFs) serve as a goodness of fit between a pair of distributions. The $d$ parameter varies between 0 and 1, the former indicates the best match and the latter means no agreement at all. 

\begin{figure}[!hpt]
    \centering
    \includegraphics[width=8.5cm, trim = 0 0 0 0, clip]{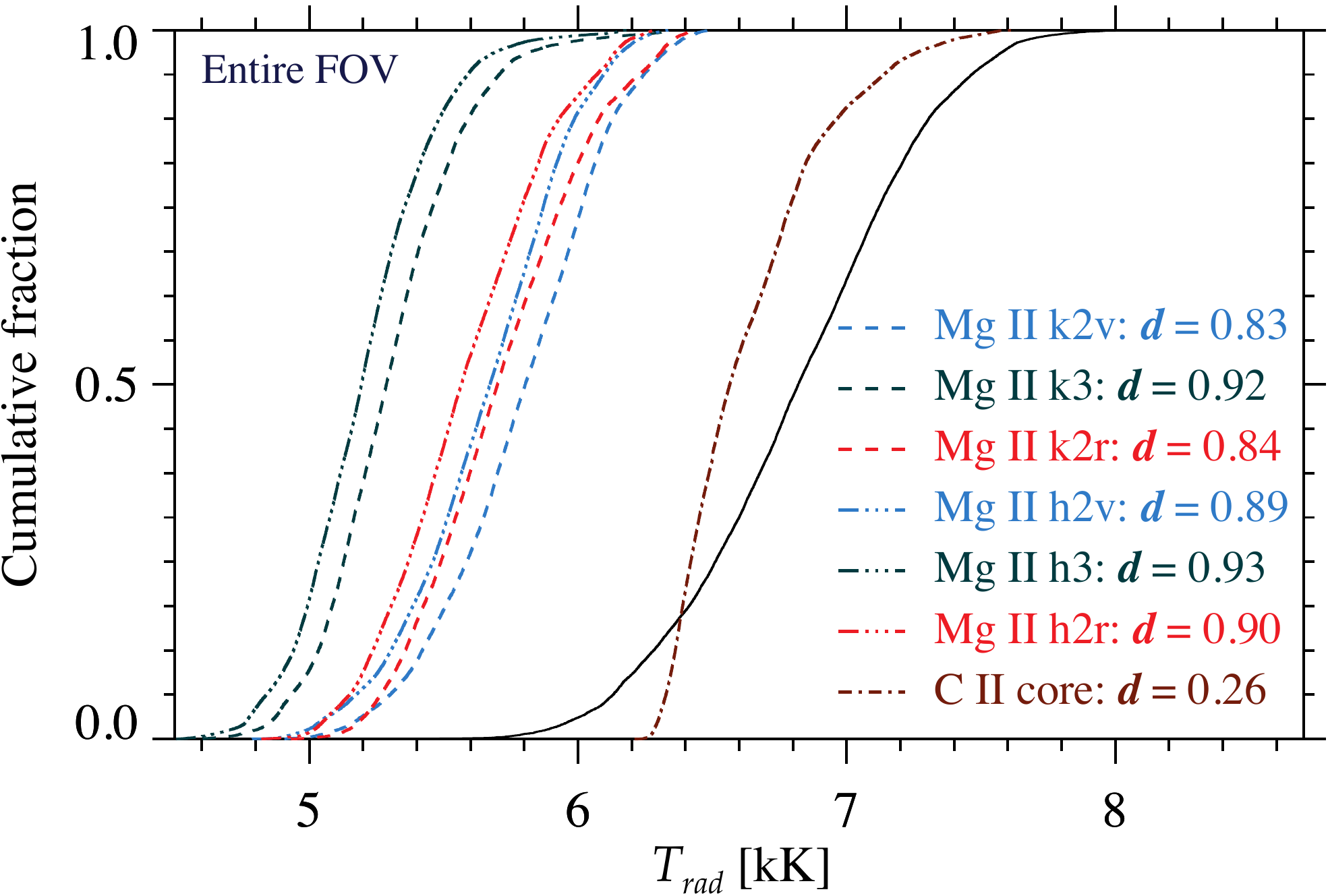}
    \caption{Cumulative distribution functions of radiation temperature of pixels with maximum sampling time difference of two minutes between the ALMA (solid line) and UV diagnostics (see the legend). The Kolmogorov–Smirnov statistics ($d$) has been provided for distribution comparison of each UV channel with ALMA.}
  \label{fig:kstest}
\end{figure}

Figure~\ref{fig:kstest} illustrates the CDFs of $T_{rad}$ distributions shown in Figure~\ref{fig:scatterplots_all}, for the ALMA (black solid line), C\,{\sc ii} (dot-dashed line) and for the blue and red emission peaks (blue and red, respectively) and line core (dark green) of the Mg\,{\sc ii}\,k (dashed lines) and of the Mg\,{\sc ii}\,h (double-dotted-dashed lines). The $d$ values, as a result of the K-S test, are printed in the figure. It is evident that only $T_{rad}$ distribution of the C\,{\sc ii} line has, to some extent, agreement with that of the ALMA 1.3~mm.
However, we note that although the radiation temperatures may be different between Mg\,{\sc ii} and ALMA~1.3~mm maps (i.e. the offsets between their CDFs), their distributions are reasonably correlated when the one-to-one comparison was performed (see the correlation coefficients in Figures~\ref{fig:scatterplots_all}-\ref{fig:scatterplots_qs}; also the comparison in Figure~\ref{fig:correlations_comp}). The offsets between their radiation temperatures can be due to the non-LTE formation of the Mg\,{\sc ii} lines (i.e. the source function is somewhat decoupled from the Planck function), resulting in different (lower) radiation temperatures compared to their actual gas temperatures (whereas the ALMA brightness temperatures represent, to a large extent, the actual gas temperatures). Hence, we cannot rule out (nor confirm) possible agreements between the actual temperature distributions that would be obtained from observations of the Mg\,{\sc ii} diagnostics and of the ALMA Band~6 (particularly, because there are good correlations between their $T_{rad}$ distributions).

A better agreement between the $T_{rad}$ distributions of the ALMA and C\,{\sc ii} line core has been obtained. In addition, we have found relatively good correlations between their one-to-one comparison (see Figure~\ref{fig:correlations_comp}). However, a $T_{gas}=2T_{rad}$ relationship was found by \citet{Rathore2015a} from a numerical simulation. Thus, the degree of agreements and/or correlations between the actual gas temperatures of the C\,{\sc ii} line-core and of the ALMA~1.3~mm remains unclear.

For simplicity, we only showed here the K-S test for pixels from all magnetic fields regions together. A very similar result was obtained for the individual regions of interests studied in Section~\ref{subsec-scatterplots}. The only difference was the shifts in temperatures while the K-S test resulted in similar parameters as shown in Figure~\ref{fig:kstest}.

\section{Summary and conclusions}
\label{sec-conclusion}

Our prime emphasis in this paper has been to present the distributions of radiation temperature in the solar chromosphere inferred from observations at millimetre and ultraviolet wavelengths with ALMA and IRIS. In particular, we have shown the radiation temperature distributions in various solar regions with different levels of underlying magnetic fields.

It was revealed that spectral splitting of the ALMA observations in a particular Band (here, in Band 6) results in a slightly higher spatial resolution. Otherwise, the standard spectral averaging smears out small-scale structures, which as a result, biases, to some extent, the temperature distributions.

The ALMA Band 6 represents a mean brightness temperature of $5.6\pm0.4$~kK in the umbra (with the largest amount of magnetic fields compared to the other solar regions under study). That is $6.5\pm0.4$~kK in the quasi-quiet regions in the immediate vicinity of the sunspot. The ALMA's band 6 poses its highest brightness temperature in plage areas (with a mean value of $7.0\pm0.4$~kK).

We found that in all solar regions the IRIS magnesium lines represent the lowest mean radiation temperature in the order of 4.2--5.7~kK within the lines under study, with the magnesium line-cores the lowest.

The brightness temperature measured with the ALMA observations at band 6 (i.e. at 1.3~mm) is, on average, compatible with that retrieved from the line core of the C\,{\sc ii}, with an exception in the umbra, where the latter is, on average, hotter by 1~kK. This may suggest a very different formation heights between the non-LTE C\,{\sc ii} line and the LTE ALMA band in the umbra, compared to the other regions under study.

The lower mean radiation temperatures of all the UV and millimetre bands in the umbra compared to other regions under study may in part be caused by Wilson depression \citep{Bray1964} causing lower formation heights, due to the larger magnetic fields (hence, larger spectral-line weakening; lower density and /or opacity).
Also, it has been shown that the sunspot atmospheres are simply cooler compared to the other regions because of reduced local heating in there \citep[e.g.][]{Maltby1986}. In plage areas, we found the highest radiation temperatures in all spectral channels. In particular, these radiation temperatures are considerably larger than the quasi-quiet areas for the magnesium lines and the ALMA 1.3~mm. The excess brightness of the magnesium lines in plage can be associated with the enhanced emission peaks at the presence of magnetic field which results in an elevated formation height compared to the average quiet-Sun (similar to the discussion on Ca\,{\sc ii}\,H line by \citealt{Jafarzadeh2017b}). 

We further examined the pixel-to-pixel relationships between radiation temperatures of the various UV bands (as well as intensities of two optically-thin lines) from IRIS and the brightness temperature of the millimetre band from ALMA. Only pixels with a maximum observing-time difference of 2 and 0.5 minutes between the two observations were included in two similar analyses. We found a good overall (direct) correlations between most of the UV channels and the millimetre band in various solar regions. This suggests that the spatial variation of the UV radiation temperatures (in a particular region of interest under study) is similar to that of the ALMA 1.3~mm. The relationships between the far-UV channels and the ALMA 1.3~mm were found to be non-linear.

Despite the (good) degree of correlations between the radiation temperatures of the ALMA Band 6 and the UV diagnostics, there is always an offset between the $T_{rad}-T_{rad}$ trend of the scatter points and the $y=x$ line (the exception is when the C\,{\sc ii} line-core is compared to ALMA; see  below). The offsets are likely due to the non-LTE formation of the UV channels, which in turn, causes the source function to be different from the Planck function. Hence, the radiation temperatures do not represent the gas temperatures (whereas in the case of ALMA Band 6, the radiation temperature is representative of the gas temperature). In addition, the sampling time differences between the ALMA and IRIS scans (although it was, somewhat, taken into account) as well as (possibly) the different heights of formation of the various spectral channels, could contribute, to some degree, in the offsets, depending on the spectral line. In particular, different formation heights should be kept in mind when the distributions of radiation temperature of the IRIS C\,{\sc ii} and ALMA 1.3~mm are compared (in addition to the non-LTE effect of the C\,{\sc ii} line), even if there is only a small offset between the linear trends and the $y=x$ line in their scatter plots.

\citet{Bastian2017} also compared the radiation temperatures from ALMA to those from a combination of the average of the h2v and h2r emission peaks with the singly peaked spectra (of the Mg\,{\sc ii}\,h line). In the present work, we studied the associations separately for each emission peaks and the line core. Also, the singly peaked spectra were interpreted as line core intensities in the present work. 
The linear correlation coefficients reported by \citet{Bastian2017,Bastian2018} for three different regions of umbra, plage and quiet regions are, respectively, higher by a factor of 8.5, 1.04, and 1.03 compared to those found here. We note that the two studies have used different line parameters, slightly different FOV, different conventional approach to define the magnetic regions, and consideration of the sampling time when pixels in the two maps are compared. The consideration of the time differences is particularly essential in the umbra where solar scene changes in time scales shorter than 2 minutes occur (e.g. umbral flashes). How these differences could result in different correlations is, however, unclear. The very small correlation coefficients we found in the umbra could partly be due to a large uncertainty introduced by the very few data points entering the analysis (because of the consideration of small time differences).

To summarize, we have provided radiation temperatures in various magnetic regions in the solar chromosphere. Relatively good correlations were found between the $T_{rad}$ of various UV diagnostics from IRIS and that of the millimetre band from ALMA. The only exception is in the umbra where fewer data points entered the analysis. In addition, we expect a larger variation in temperature with height in the umbra (due to, e.g. sunspot waves). The comparison suggested that the ALMA Band 6 and the various IRIS diagnostics provide information about the heating events in likely different chromospheric heights; ALMA represents the actual gas temperature, the UV channels the radiation temperatures (which are, to some extent, lower than the gas temperatures).

Inspection of similar solar regions from the synthetic UV and millimetre bands studied here, as well as their characteristic analysis, could clarify the relationships between radiation and gas temperatures of the different spectral channels and between their heights of formation. Future work will aim at such investigations.

\begin{acknowledgements}

This research has received funding from the European Research Council (ERC) under the European Union’s Horizon 2020 research and innovation programme (grant agreement No. 682462) and is supported by the Research Council of Norway through its Centres of Excellence scheme, project number 262622.
This paper makes use of the following ALMA data: ADS/JAO.ALMA\#2011.0.00020.SV. ALMA is a partnership of ESO (representing its member states), NSF (USA) and NINS (Japan), together with NRC (Canada), NSC and ASIAA (Taiwan), and KASI (Republic of Korea), in cooperation with the Republic of Chile. The Joint ALMA Observatory is operated by ESO, AUI/NRAO and NAOJ. IRIS is a NASA small explorer mission developed and operated by LMSAL with mission operations executed at NASA Ames Research center and major contributions to downlink communications funded by ESA and the Norwegian Space Centre. The AIA and HMI data are courtesy of the NASA/SDO, as well as AIA and HMI science teams. SJ is grateful to L.~Rouppe van der Voort, S.K.~Solanki, T.M.D.~Pereira, and L.~Mas-Ribas for useful discussions.

\end{acknowledgements}

\bibliographystyle{aa}
\bibliography{Shahin_ALMA_IRIS_comp}

\end{document}